\documentclass[showpacs,preprintnumbers,amsmath,amssymb,nofootinbib,superscriptaddress,twocolumn]{revtex4}

\usepackage{graphicx}
\usepackage{amsmath}
 \usepackage{color}

\begin{document}

\title{Scaling laws for diffusion on (trans)fractal scale-free networks}

\author{Junhao Peng}
\affiliation{
  School of Math and Information Science, Guangzhou University, Guangzhou 510006, China.}
\affiliation{
  Key Laboratory of Mathematics and Interdisciplinary Sciences of Guangdong Higher Education Institutes, Guangzhou University, Guangzhou 510006, China.}

\author{Elena Agliari}
\affiliation{
Department of Mathematics, Sapienza Universit\`a di Roma, 00198 Rome, Italy.}
\affiliation{
Istituto Nazionale di Alta Matematica, Rome, Italy.}

\begin{abstract}
Fractal (or transfractal) features are common in real-life networks and are known to influence the dynamic processes taking place in the network itself. Here we consider a class of scale-free deterministic networks, called $(u,v)$-flowers, whose topological properties can be controlled by tuning the parameters $u$ and $v$; in particular, for  $u>1$, they are fractals endowed with a fractal dimension $d_f$, while for $u=1$, they are transfractal endowed with a transfractal dimension $\tilde{d}_f$. In this work we investigate dynamic processes (i.e., random walks) and topological properties (i.e., the Laplacian spectrum) and we show that, under proper conditions, the same scalings (ruled by the related dimensions), emerge for both fractal and transfractal.
 \end{abstract}

\pacs{05.40.Fb, 05.45.Df, 05.60.Cd}

\maketitle

\noindent
\textbf{
Many real-life networks are known to exhibit a fractal-like topology with a power-law degree distribution and these features strongly affect the dynamic processes possibly taking place in the network itself. Graph models that display such properties allow mathematical investigations towards a quantitative understanding of the role of topology in transport efficiency. In this context, $(u,v)$-flowers provide a very useful model as they constitute a class of deterministic networks which, by properly tuning the parameters $u$ and $v$, span a wide range of topological properties and are feasible for exact analytical treatment through renormalization techniques.
Here, we consider $(u,v)$-flowers and we focus on the first-passage time (FPT) for random walks (which is a key indicator of how fast information, mass, or energy diffuse in a given system) and on the Laplacian spectrum (which crucially controls dynamic properties such as heat transport and mixing time). As for the random walk problem, through a generating function formalism, we are able to rigorously obtain exact results for the FPT and related quantities, which are expressed in terms of the network size. As for the Laplacian spectrum, we are able to highlight proper scalings for the eigenvalue distribution and for the spectral gap versus the network size.
The robustness of these results with respect to the parameters $u$ and $v$ is also discussed. This is a rather subtle point given that, when $u=1$, fractality is replaced by transfractality and propagation ceases to be compact.
These findings pave the way to further investigations on the robustness of scalings in the presence of some degree of noise, e.g., modeled in terms of stochastic weights on links.}

\section{Introduction}
\label{intro}
Graph models feasible for an exact treatment constitute interesting examples as one can highlight analytical relations between the behaviour of processes embedded in the graph and its topology \cite{Newman, Barrat, Albert02}.
In this context, much attention has been paid to deterministic scale-free graphs, given the widespreadness of the power-law degree distribution in real-life systems (see e.g., \cite{Vicsek01,Bollt05, TeBeVo09, AgBu09}). One of the first examples of a deterministic scale-free graph was proposed in the late 70s by Berker and Ostlund \cite{Berker}. Since the late 90s, many more examples were introduced, among which we mention the model by Song, Havlin and Makse \cite{Song05,Song06} and the model by Rozenfeld, Havlin and ben-Avraham \cite{RoHa07,Enc1}, also called $(u,v)$-flowers; in this paper we focus on the later.
Actually, $(u,v)$-flowers constitute a \emph{class} of deterministic scale-free networks which, by suitably varying the parameters $u$ and $v$, can exhibit different topological properties (e.g., fractality, clustering, assortativity). Moreover, they are modular and amenable to exact analysis by renormalization techniques. All these features justify their popularity.

In fact, for the $(u,v)$-flowers many results (both analytical and numerical) have already been collected (see e.g., \cite{RoHa07,Enc1,ZhangXie09, Hwang10, zhangLiu11,Zhang11,MeAgBeVo12}).
In particular, it was proved that there are strong differences between the case $u>1$ and the case $u=1$. For  $u>1$, they are \textquotedblleft large-world\textquotedblright and fractal; for $u=1$, they are ``small-world'' and infinite dimensional (i.e., the number of nodes increases faster than any power of their diameter), yet a transfinite fractal dimension exists. Thus, in order to characterize how mass scales with its length scale, for the former ($u>1$), one defines a fractal dimension $d_f$, while, for the latter ($u=1$), one introduces a transfractal dimension\footnote{Briefly, denoting with $L$ the diameter of the network and with $N$ its volume, one has the following scalings: $N(bL) = b^{d_f} N(L)$, when $u>1$, and $N(L + b) = e^{b \tilde{d}_f} N(L)$, when $u=1$; here $b>0$ is an arbitrary factor.} $\tilde{d}_f$ \cite{Enc1}.

Beyond the fractal dimension, there are other parameters that characterize a graph (e.g., see \cite{Mieghen-2011,Enc2,Donetti-JStat2004,Agl16}). In particular, the spectral dimension $d_s$ plays a central role in characterizing dynamic quantities: for instance, the mean-square displacement $\langle r^2 \rangle$ of a random walker on a fractal scales with time $t$ as $\langle r^2 \rangle \propto t^{d_s/d_f}$. Moreover, for scale-free renormalizable graphs, one can introduce the exponent $d_k$ that characterizes the scaling transformation in the degree distribution in such a way that $\gamma = 1 + d_f/d_k$, where $\gamma$ is the exponent for the distribution.
The possibility to extend these features to the case of transfinite fractals has been only partially addressed; for instance, for scale-free transfinite fractals the natural extension $\gamma = 1 + \tilde{d}_f/\tilde{d}_k$ holds, where $\tilde{d}_k$ is the transfinite exponent analogous to $d_k$ \cite{Song05, Enc1}.
In this work we investigate the connection between the spectral dimension, a number of first passage time problems and the distribution of the Laplacian spectrum \cite{Mieghen-2011,Agl16}.

More precisely, as for random walks, the quantities we are interested in are  the  first-passage time (FPT), which is the time taken by a random walker  to reach a given site for the first time, the first-return time (FRT), which is the time taken by a random walker to return to the starting site for the first time~\cite{Redner07,  MeyChe11, Condamin05, CondaBe07, BeChKl10,SiBoni15, EiKaBu07,MoDa09,SaKa08,Agl16}, and the global first-passage time (GFPT), which is the first-passage time  from a randomly selected site to a given site~\cite{HvKa12}. In the past years, significant work has been devoted to analyzing these quantities on different kinds of networks (see e.g.,~\cite{Avraham_Havlin04,Redner07,CondaBe07,Metzler-2014}).
Remarkably, the mean and the variance of these quantities are intrinsically related to the spectral dimension (whenever defined) of the underlying graph \cite{CondaBe07,BeChKl10,HvKa12}: on some special fractal lattices, Haynes \emph{et al} found that the mean of the FPT scales with the volume $N$ of network as $N^{2/d_s}$ and the variance of the FPT scales with the volume of network as $N^{4/d_s}$~\cite{HaRo08}; on scale-free networks, Hwang \emph{et al} found that the mean of the FPT is affected by the spectral dimension and by the exponent of the degree distribution~\cite{HvKa12}.

As for the Laplacian spectrum, we are interested in the eigenvalue distribution $P(\lambda)$, which, in the small eigenvalue limit ($\lambda \rightarrow 0$), scales as $P(\lambda) \sim \lambda^{d_s/2-1}$ \cite{Orbach, Rammal}.
The spectral density has been shown to control the dynamic properties of homogeneous or fractal systems such as heat transport, specific heats, scattering of waves, etc. \cite{Mieghen-2011,Enc2}. Given such a wide range of applicability, the study of Laplacian eigenvalues and eigenvectors has been the subject of extensive investigations. In particular, the full knowledge of the spectrum has been achieved for several deterministic structures among which we can mention several examples of fractals \cite{Cosenza-PRA1992,Jayanthi-PRL1992}, of small-world networks \cite{Zhang-SciRep2015}, and of scale-free graphs \cite{Zhang-PRE2013b,Agliari-SciRep2017}.

For the $(u,v)$-flowers considered here, we expect that, as long as $u>$1, we can recover the above mentioned scalings ruled by the fractal dimension and by the spectral dimensions, whose values  $d_f = \log (u + v) / \log(u)$ and $d_s = 2 \log(u+v)/ \log (uv)$, respectively, have been previously found \cite{RoHa07,Enc1}. When $u=1$, the scenario is more subtle, yet we will show that it is possible to extend the previous formula as $\tilde{d}_s = 2 \log (1+v) / \log(v)$, through which we recover the expected scalings for a first-passage problem and for the Laplacian spectrum.

This paper is structured as follows. In Sec.~\ref{sec:TF} we describe the iterative building procedure to get the $(u,v)$-flowers and we review their main topological properties. Then, in Sec.~\ref{sec:FP}, we analyze first passage quantities for random walks embedded in $(u,v)$-flowers; exploiting the generation function formalism, we are able to get, for arbitrary $u$ and $v$, exact results for the mean and the variance of FPT, FRT and GFPT. These will also be corroborated via numerical simulations. Next, in Sec.~\ref{sec:laplacian} we analyze numerically the Laplacian spectrum and its distribution. Finally, Sec.~\ref{sec:4} is left for conclusions and discussions. Technicalities on calculations are all collected in the Appendices.

\section{Network model}
\label{sec:TF}

\begin{figure}
\begin{center}
\includegraphics[scale=0.55]{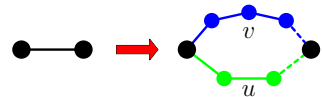}
\caption{Iterative construction of the $(u,v)$ flowers. The $(u,v)$ flower of generation $k$ $(k>0)$, denoted by $G(k)$,  is obtained from $G(k-1)$ by replacing every edge of $G(k-1)$ by  two parallel paths with lengths $u$  and $v$ ($1\leq u \leq v$), as shown on the right-hand side of the arrow.}
\label{fig:1}       
\end{center}
\end{figure}

\begin{figure}
\begin{center}
\includegraphics[scale=0.55]{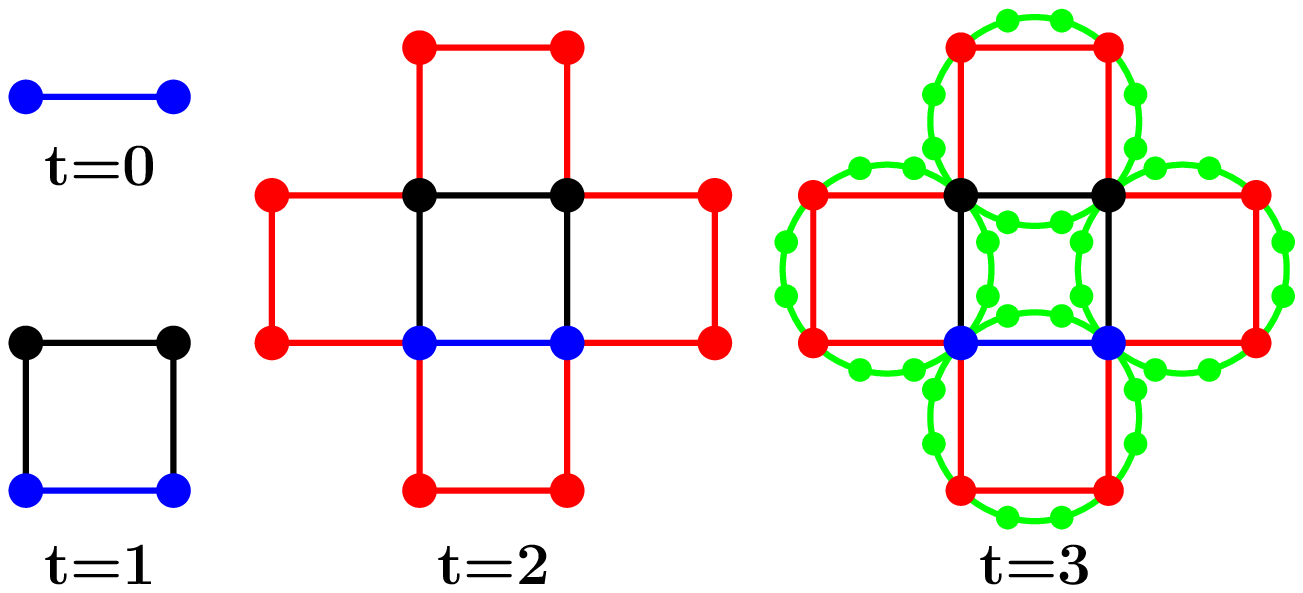}
\caption{The construction of the $(1,3)$-flowers with generations $t=0, 1, 2, 3$. }
\label{fig:2}       
\end{center}
\end{figure}

\begin{figure}
\begin{center}
\includegraphics[scale=0.55]{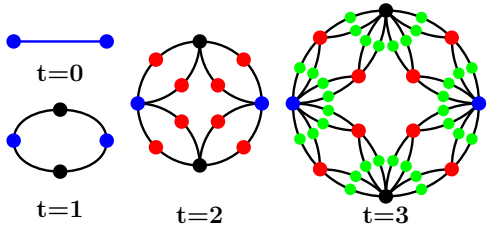}
\caption{The construction of the $(2,2)$-flowers with generations $t=0, 1, 2, 3$. }
\label{fig:3}       
\end{center}
\end{figure}

The networks considered here, also called $(u,v)$-flowers ($1\leq u\leq v$), are deterministic and recursively grown~\cite{RoHa07}.  Let $G(t)$ denote the $(u,v)$-flower of generation $t$ ($t \geq 0$).  The construction starts from two nodes connected by an edge, which corresponds to $G(0)$.  For $t \geq 1$, $G(t)$ is obtained  by replacing every edges of $G(t-1)$  by two parallel paths of length $u$ and $v$, respectively\footnote{Notice that, without loss of generality, one can focus on the case $u \leq v$.}.
Fig.~\ref{fig:1} shows the iterative building procedure and  Figs.~\ref{fig:2}-\ref{fig:3}
show the construction of the $(1,3)$-flower and of the $(2,2)$-flower, respectively, for generations $t=0, 1, 2, 3$.
Letting $w=u+v$, one can see that for $G(t)$ the total number of edges is $E_t = w^{t}$ and the total number of nodes is  $N_t=[(w-2)w^{t}+w]/(w-1)$~\cite{RoHa07,Zhang11}. Among the whole set of nodes, we distinguish $w$ main hubs (i.e., nodes with the highest degree), corresponding to the ``ancestors'', namely the original nodes of $G(1)$, and whose degree is $2^t$.

%

\begin{figure}
\begin{center}
\includegraphics[scale=0.6]{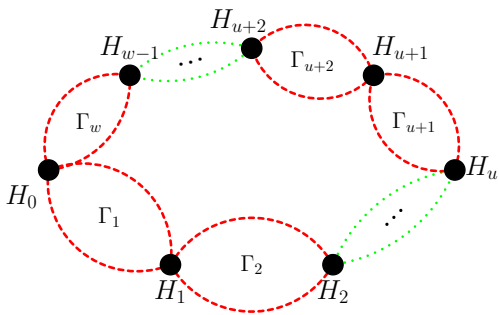}
\caption{Alternative construction of the $(u,v)$-flower which highlights self-similarity: the network of generation $t$, denoted by $G(t)$, is composed of $w\equiv u+v$ copies, called subunits, of $G(t-1)$ which are labeled as  $\Gamma_1$,  $\Gamma_2$,  $\cdots$ ,$\Gamma_w$, and connected to one another at its $w$ hubs, denoted by $H_0$, $H_1$,$\cdots$, $H_{w-1}$.
}
\label{Self_similar}
\end{center}
\end{figure}

The network also has an equivalent  construction method which highlights its self-similarity~\cite{RoHa07}.
Referring to Fig.~\ref{Self_similar}, in order to obtain $G(t)$, one can make $w$ copies of $G(t-1)$ and join them  at their hubs denoted by $H_0$, $H_1$,$\cdots$, $H_{w-1}$. In this way, $G(t)$ is composed of $w$ copies of $G(t-1)$ labeled as $\Gamma_1$, $\Gamma_2$, $\cdots$, $\Gamma_w$, which are connected with each other by the $w$ hubs.
Notice that the self-similarity of these networks, along with the fact that different replicas meet at a single node, makes them amenable to exact analysis by renormalization techniques.

We now review the main topological properties of $(u,v)$-flowers, while we refer to \cite{RoHa07,Enc1} for a more extensive treatment.
\newline
The degree distribution $P(k)$ of $(u,v)$-flowers follows a power-law $P(k)\sim k^{-\gamma}$  with the exponent $\gamma=1+\log(u+v)/\log 2$. This can be obtained, by construction, since in these networks nodes can only have degree $k = 2^m$, with $m=1, ..., t$.
Moreover, as anticipated in the introduction, for  $u>1$, the networks are fractal with fractal dimension $d_f= \log (u+v)/\log u$~\cite{RoHa07}, walk dimension $d_w= \log (u v)/ \log u$, and spectral dimension $d_s=2 d_f /d_w = \log(u+v)/ \log (uv)$~\cite{Hwang10}. For $u=1$, the networks have infinite dimension, infinite walk dimension but finite spectral dimension $d_s$~\cite{Hwang10}. Indeed, the mass of the networks and the time for diffusion between any two nodes increase faster than any power (dimension) of their diameter.
 However one can still characterize the scaling of the mass and of the diffusion time by noticing that $(1,v)$-flowers are transfractals. This term refers to networks that are self-similar and have infinite dimension, but finite transfinite fractal (or transfractal, for short) dimension $\tilde{d}_f$. The latter is defined as follows: being $N(L)$ the number of nodes falling within a (chemical) distance $L$ from a given arbitrary node, then upon an additive mapping $L \rightarrow L + b$, we get $N(L+b) = e^{b \tilde{d}_f}N(L)$. Similarly, being $T(L)$ the characteristic time for diffusion between any two nodes at distance $L$, one has $T(L+b) = e^{b \tilde{d}_w}T(L)$, where $\tilde{d}_w$ is the transfinite exponent analogous to the walk dimension $d_w$ and is referred to as transwalk dimension \cite{Song06,RoHa07}.
For $(1,v)$-flowers,
 $\tilde{d}_f=\log(1+v)/(v-1)$ and $\tilde{d}_w=\log(v)/(v-1)$ \cite{RoHa07}. By exploiting the relation  $d_s=2 d_f /d_w$ \cite{Rammal}, we can naturally define a spectral dimension as $\tilde{d}_s=2\tilde{d}_f/\tilde{d}_w=2 \log (1+v)/ \log (v)$. Notice that $\tilde{d}_s$ recovers the expression given above for $d_s$ as $u$ is set equal to $1$, that is, the spectral dimension does not exhibit any singularity as long as $u$ and $v$ are not \emph{both} equal to $1$. On the other hand, the case $u=v=1$ is a dimer connected by $2^t$ links.

Finally, we notice that, even though $d_s$ can be trivially extended from the fractal to the transfractal regime, the two regimes exhibit different diffusive behaviors. In fact, when $u,v>1$ one has $d_s<2$ and the exploration is compact, while when $u=1$ (and $v>1$) one has $\tilde{d}_s>2$ and the likelihood for the random walker to visit a certain node is expected to depend more sensitively on the kind of node (either hub or simple node) \cite{HvKa12}.

\section{First-passage problems on $(u,v)$-flowers} \label{sec:FP}
The first-passage problem on a $(1,2)$-flower has been extensively investigated in the past few years~\cite{Hwang10, zhang10, ZhangLin15,  peng15, PengAgliariZhang15}; the mean of the FPT  to a given hub is also obtained for some special cases, such as $(u=2, v=2)$ and $(u=1, v=3)$ ~\cite{ZhangXie09, MeAgBeVo12}.
However, for generic $u$ and $v$, the exact expressions for FPT, FRT, and GFPT at finite size $N_t$ are still unknown and will be obtained in this paper. From these expressions we will derive scalings with respect to $N_t$ that hold for relatively large networks.
\newline
Before proceeding, it is worth noticing that these first-passage quantities are affected by the position of the source and of the target. Here we are not going to enumerate and to analyze all the possible cases but we shall consider the FPT from one given hub to another given hub, and the FRT and the GFPT to a given hub. Also, when evaluating the GFPT, the starting site is selected by mimicking the  steady state, namely the probability that a node $v$ is selected as starting site is $d_v/(2E_t)$,  where $E_t$ is the total number of edges of the network and $d_v$ is the degree of node $v$.
\newline
Moreover, the analytical calculations presented in the following subsections use probability generating function theory.
Technicalities on this tool can be found in Appendix \ref{sec:PGF}, while here we briefly review the definition (see e.g., \cite{Gut05}).
Let $p_k$ ($k=0,1,2,\cdots$) be the probability mass function of a discrete random variable $T$ taking values in the non-negative integers $\{0,1, ...\}$, then the related probability-generating function $\Phi_{T}(z)$ of $p_k$ is
 \begin{equation}\label{Def_PGF}
   \Phi_{T}(z)=\sum_{k=0}^{+\infty}z^k p_k.
 \end{equation}

\subsection{Probability generating function and the first, second moment of the FPT from $H_0$ to $H_u$}

Let $F_{H_0\rightarrow H_u}(t)$ be the FPT from two arbitrary hubs $H_0$ to $H_u$ on the network $G(t)$ and let $\Phi_{FPT}(t,z)$ be the related probability generating function.
In this subsection we first obtain a recursive equation for $\Phi_{FPT}(t,z)$ and, from this, we get explicit expressions for the first and the second moment of $F_{H_0\rightarrow H_u}(t)$.

We recall that, for $t=0$, the network is constituted by two nodes (i.e., $H_0$ and $H_u$) connected by an edge, and one has $\Phi_{FPT}(0,z)=z$.
For $t=1$, the network is constituted by a ring with $w=u+v$ nodes and, as explained in the Appendix \ref{PGF_FPTg1}, we can build linear equations of probability generating functions based on the relation among transferring probabilities. In fact, we find
\begin{equation}
\label{PGF_FPT_G1}
  \Phi_{FPT}(1,z)= \left(\frac{z}{2} \right)^u\frac{\delta_{w-u-1}}{\delta_{w-1}}+\sum_{i=w-u}^{w-1} \left(\frac{z}{2} \right)^{2i-w+u}\frac{\delta_{w-u-1}}{\delta_{i}\delta_{i-1}},
\end{equation}
where
\begin{equation}
\label{delta_i}
  \delta_i=c_1\times \left( \frac{1+\sqrt{1-z^2}}{2} \right)^i+c_2\times \left(\frac{1-\sqrt{1-z^2}}{2} \right)^i,
\end{equation}
with $c_1=\frac{1-z^2+\sqrt{1-z^2}}{2(1-z^2)}$ and $c_2=\frac{1-z^2-\sqrt{1-z^2}}{2(1-z^2)}$.

For  any $t>1$, according to the self-similar structure of the network,  we can construct $L$ independent random variables $T_1$, $T_2$, $\cdots$, such that
\begin{equation}
\label{FPT_pathlength}
  F_{H_0\rightarrow H_u}(t)= T_1+T_2+\cdots+T_L,
\end{equation}
where $L$ is the FPT  from $H_0$ to $H_u$  on the  $(u,v)$-flower of generation $1$ and $T_i$ ($i=1,2,\cdots$) are identically distributed random variables, each of them can be regarded as the first-passage time from $H_0$ to $H_u$  on $G(t-1)$.
Thus, for any $t>1$, one can show that the $F_{H_0\rightarrow H_u}(t)$ probability generating function satisfies
\begin{equation}\label{Rec_PGF_FPTm}
  \Phi_{FPT}(t,z)=\Phi_{FPT}(1,\Phi_{FPT}(t-1,z)).
\end{equation}
By taking the first and second order derivative on both sides of Eq.~(\ref{Rec_PGF_FPTm}) and posing $z=1$,
we obtain the mean $\langle FPT_t\rangle$ and the second moment $\langle FPT^2_t\rangle$ of $F_{H_0\rightarrow H_u}(t)$:
\begin{eqnarray}
\label{MFPT_Em}
    \langle FPT_t \rangle &=& (uv)^{t}, \\
\label{M2FPTm}
   \langle FPT^2_t\rangle&=&(u^2\!+\!3uv\!+\!v^2\!-\!5)\frac{(uv)^{2t}\!-\!(uv)^{t}}{3(uv-1)}\!+\!(uv)^{t}.
\end{eqnarray}
Therefore, the variance $\textrm{Var}(FPT_t) \equiv  \langle FPT^2_t\rangle - \langle FPT_t\rangle^2$ of the FPT is
\begin{equation}
\label{R_Second_m_FPT}
\textrm{Var}(FPT_t) =  \frac{u^2+v^2-2}{3(uv-1) } \left[ (uv)^{2t}-(uv)^t\right].
\end{equation}
The derivation of Eqs.~(\ref{Rec_PGF_FPTm}), (\ref{MFPT_Em}) and (\ref{M2FPTm})  are presented in the Appendix~\ref{App:PGF_FPTt}.

\subsection{Recursive equations for GFPT and FRT probability generating functions and related moments}\label{PGF_RT_GFPT}
In this subsection we present recurrence relations for the GFPT and FRT probability generating functions for a given hub. Calculations are rather lengthy and, again, in the main text we provide only the final results, while more details can be found in the Appendices \ref{App:RT}-\ref{sec:Moments1}. 

We analyze the GFPT and the FRT for the arbitrary hub $H_0$.
Let $P_{i\rightarrow j}(t,k)$ ($k=0,1,2,\cdots$) be the probability distribution of the FPT from site $i$ to site $j$ on the network $G(t)$. In particular, $P_{H_0\rightarrow H_0}(t, k)$ ($k=0,1,2,\cdots$) are the probability distributions of the FRT to $H_0$ on the network $G(t)$.
The probability distribution for the GFPT to hub $H_0$ on $G(t)$, denoted as $P_{H_0}(t,k)$ ($k=0,1,2,\cdots$), can be represented by
\begin{equation}\label{Def_GFPP}
  P_{H_0}(t,k)=\sum_{i}\frac{d_i}{2E_t}P_{i\rightarrow H_0}(t,k),
\end{equation}
where the sum runs over all the nodes of $G(t)$.

The generating functions of $P_{H_0\rightarrow H_0}(t, k)$ ($k=0,1,2,\cdots$) and of $P_{H_0}(t,k)$ ($k=0,1,2,\cdots$) are denoted as  $\Phi_{FRT}(t,z)$ and $\Phi_{GFPT}(t,z)$, respectively. Otherwise stated, these are, respectively, the FRT and GFPT probability generating functions for the hub $H_0$ (see Eq.~(\ref{Def_PGF})). In order to get an expression for $\Phi_{FRT}(t,z)$ and $\Phi_{GFPT}(t,z)$ we first need to analyze the probability generating function of return time (not necessarily for the first time) to the node $H_0$, whose generating function is denoted as $\Phi_{RT}(t,z)$. In fact, it is known that~\cite{HvKa12},  
\begin{equation}\label{R_GFPT_RT}
   \Phi_{GFPT}(t,z)=\frac{2^{t-1} z}{(1-z)w^{t}}\times\frac{1}{ \Phi_{RT}(t,z)},
\end{equation}
and
\begin{equation}\label{R_FRT_RT}
 \Phi_{FRT}(t,z)=1-\frac{1}{\Phi_{RT}(t,z)}.
\end{equation}
Now, letting $\psi(z)\equiv\frac{\Phi_{RT}(0,z)}{\Phi_{RT}(1,z)}$ (see Eq.~(\ref{psiz})) and plugging the expression for $\Phi_{RT}(t,z)$ found in Appendix \ref{App:RT} (see Eq. (\ref{R_RT})) into  Eqs.~(\ref{R_FRT_RT}) and (\ref{R_GFPT_RT}), we get
\begin{equation}\label{R_GFPTm}
  \Phi_{GFPT}(t,z)=\frac{2}{u+v}\psi(\Phi_{FPT}(t-1,z))\Phi_{GFPT}(t-1,z),
\end{equation}
and
\begin{equation}\label{R_FRTm}
  \Phi_{FRT}(t,z)=1-\psi(\Phi_{FPT}(t-1,z))[1- \Phi_{FRT}(t-1,z)],
\end{equation}
with initial conditions $\Phi_{GFPT}(0,z)=\frac{1}{2}(z+1)z$ and $\Phi_{FRT}(0,z)=z^2$. 
\newline

Next, we obtain  exactly  the first and the second moment of the GFPT,
by calculating the first and the second order derivatives  on both sides of Eq.~(\ref{R_GFPTm}) and posing  $z=1$. As explained in Appendix \ref{sec:Moments2}, we get
\begin{eqnarray}\label{R_First_m_GFPTm}
 \langle GFPT_t \rangle  &=& \frac{3}{2}+\left[\frac{(u+v)^2}{6}-\frac{2}{3}\right]\frac{(uv)^t-1}{uv-1}, \\
 \label{R_Second_m_GFPTm}
\langle GFPT_t^2 \rangle & = &  \frac{5}{2}\! +\!k_1\frac{(uv)^{2t}\!-\!1}{(uv)^2-1}\! +\!k_2\frac{(uv)^t\! -\!1}{uv-1},
\end{eqnarray}
where $k_1$ and $k_2$ are proper functions of $u$ and $v$ (see Eqs.~(\ref{K_1})-(\ref{K_2})). Finally, for large networks sizes, combining the previous expressions (and taking care of the smaller orders as reported in Appendix \ref{sec:Moments2}), we get
\begin{equation}
   \label{VGFPT}
\textrm{Var}(GFPT_t) \sim \langle GFPT_{t} \rangle^2.
\end{equation}

Let us now move to the FRT and evaluate the related first and second moment by calculating the first and second order derivative of the related probability generating function $\Phi_{FRT}(t,z)$ in Eq.~(\ref{R_FRTm}). As explained in Appendix \ref{sec:Moments1}, we get
\begin{eqnarray}
\nonumber
\langle FRT_t \rangle &=& 2\times\left(\frac{u+v}{2}\right)^t,\\
     \nonumber
     \langle FRT^2_t\rangle &=& \left(\frac{u+v}{2}\right)^t \left \{ \frac{2}{3} \frac{(uv)^t-1}{uv-1} [(u+v)^2-4]
      + 4 \right\}.
\end{eqnarray}
Notice that, for large sizes, we can write $\langle FRT^2_t\rangle\sim \langle FRT_{t}  \rangle \langle GFPT_{t}\rangle$, and
the variance $\textrm{Var}(FRT_t)$ of the FRT scales as
  \begin{equation}
   \label{S_M2FRT}
   \textrm{Var}(FRT_t) \sim \langle FRT_{t}  \rangle \langle GFPT_{t}\rangle.
\end{equation}

\subsection{Scalings}
In this subsection we resume the results found for the first-passage quantities considered in the previous subsections and derive their scaling with the system size. 
Recalling that, in the large volume limit, $N_t = [(w-2)w^t + w]/(w-1)\sim (u+v)^t$ (see Sec.~\ref{sec:TF}), we can reshuffle Eqs.~(\ref{MFPT_Em}), (\ref{R_Second_m_FPT}), (\ref{R_First_m_GFPTm}), (\ref{VGFPT})
to obtain
\begin{eqnarray}
\label{GMFPT}
\langle FPT_t \rangle &\sim& \langle GFPT_t \rangle \sim N_t^{\frac{\log(uv)}{\log(u+v)}},\\
\label{S_Var_GFPT}
\textrm{Var}(FPT_t) &\sim& \textrm{Var}(GFPT_t) \sim N_t^{\frac{2 \log (uv)}{\log(u+v)}}.
\end{eqnarray}

For the case $u>1$, the $(u,v)$-flowers have spectral dimension $d_s=\frac{2 \log(u+v)}{\log(uv)}$~\cite{Hwang10}.  Therefore Eqs.~(\ref{GMFPT}) and (\ref{S_Var_GFPT}) imply that, $\langle GFPT_t\rangle \sim N_t^{{2}/{d_s}}$ and $\textrm{Var}(GFPT_t) \sim N_t^{{4}/{d_s}}$. As expected, both the mean and the variance of the GFPT are controlled by the spectral dimension $d_s$, and we recover the results by Haynes and Roberts \cite{HaRo08}.

For the case $u=1$, the $(u,v)$-flowers are transfractals with the transfractal dimension $\tilde{d}_f=\log(1+v)/(v-1)$ and transwalk dimension $\tilde{d}_w=\log(v)/(v-1)$~\cite{RoHa07}. Thus, the spectral dimension remains finite and is given by $\tilde{d}_s=2\tilde{d}_f/\tilde{d}_w=2 \log(1+v)/ \log (v)$.
With this definition, Eqs.~(\ref{GMFPT}) and (\ref{S_Var_GFPT})  can be recast as $\langle GFPT_t\rangle \sim N_t^{{2}/{\tilde{d}_s}}$ and $\textrm{Var}(GFPT_t) \sim N_t^{{4}/{\tilde{d}_s}}$.
Therefore, the formal scalings for the mean first-passage times considered here versus the system size are robust with respect to $u$.

Notice that Hwang et al. \cite{HvKa12} have highlighted for the GFPT to an arbitrary node $i$ on $(u,v)$-flowers different scalings for the case $u>1$ and $u=1$, that is, respectively, $\langle GFPT \rangle \sim N^{2/d_s}$, independently of the degree $k_i$ of $i$, and $\langle GFPT \rangle \sim N k_i^{-\alpha}$, where $\alpha = (1 - 2/ \tilde{d}_s) (\gamma -1)$. However, exploiting that $\gamma = 1 + \log(u+v)/ \log 2$ (see Sec.~\ref{sec:TF}), one can reshuffle the previous scaling as $\langle GFPT \rangle \sim N k_i^{- (1 -2/ \tilde{d}_s) \log_2 (1+v)}$.
As a result, as long as we are considering hub nodes with degree $k_i \sim k_{\textrm{max}} \equiv \max_i (k_i) = 2^t$, we recover the same scalings holding for $u>1$, despite the network being transfractal and the exploration non-compact.

\begin{figure}
\begin{center}
\includegraphics[scale=0.26]{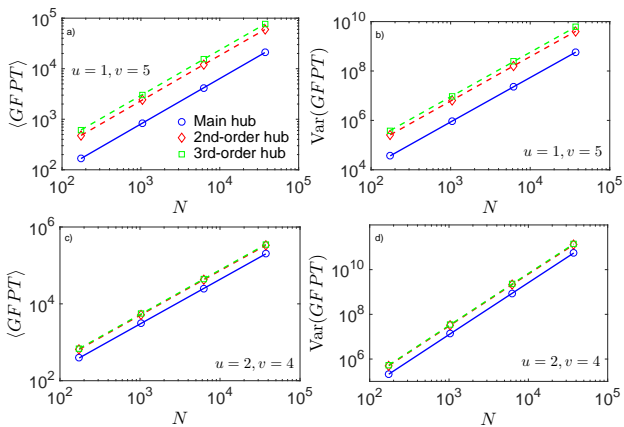}
\caption{In these panels we present in a logarithmic scale plot results for the mean (left panels) and the variance (right panels) of the GFPT versus the network size. Different target nodes (i.e., the main hub, second-order hubs and third-order hubs) are considered and shown in different colors as explained by the legend. The symbols represent results from numerical simulations; the solid lines represent the analytical results for the main hub (see Eq.~(\ref{R_First_m_GFPTm}) and Eq.~(\ref{R_Second_m_GFPTm})); the dashed lines represent the best fit according to Eqs.~(\ref{GMFPT}) and (\ref{S_Var_GFPT}). In panels $a$ and $b$ we focused on the case $u=1$ and $v=5$, while in panels $c$ and $d$ we focused on the case $u=2$ and $v=4$. In any case, the  curves for different target nodes display the same slope, consistently with Eqs.~(\ref{GMFPT}) and (\ref{S_Var_GFPT}). This picture is confirmed by other results (not shown) collected for further choices of $u$ and $v$.}
\label{fig:scalings}
\end{center}
\end{figure}

This picture is corroborated by numerical results shown in Fig.~\ref{fig:scalings}, where we considered the GFPT to a main hub with degree $2^t$, the GFPT to a ``second-order hub'' displaying the second largest degree $2^{t-1}$, and to a ``third-order hub'' displaying the third largest degree $2^{t-2}$. In all these cases, and for both $u>1$ and $u=1$, the scalings provided by Eqs.~(\ref{GMFPT}) and (\ref{S_Var_GFPT}) are recovered but, as expected, as the target degree is reduced, the mean time and its variance increase by a factor\footnote{This factor depends not only on the target degree but, more generally, on the topological centrality of the target.}. In the case $u=1$, the scaling is expected to fail when nodes with degree sublinear with $k_{\textrm{max}}$ are considered as target nodes.

As a further check, we verified numerically that the spectral dimension controls the time decay of the return-to-origin probability $R(n)$.
In fact,  for heterogeneous networks, Hwang et al. \cite{HvKa12,HvKa14} found that for intermediate times\footnote{Notice that this regime is vanishing when the target node is a hub.} $R(n) \sim n^{-{d}_s/2}$, and, accordingly, we obtained $R(n) \sim n^{\log(u+v)/ \log(uv)}$, valid for both $u>1$ and $u=1$.

\section{The Laplacian spectrum}
\label{sec:laplacian}
The Laplacian matrix has a long history in science and there are several books and survey papers dealing with its mathematical properties and applications, see e.g., \cite{Biggs-1993,Mohar-DM1992,Mieghen-2011}.
Here we review some basic definitions in order to provide a background for the following analysis.

Let $\textbf{A}$ be the adjacency matrix of an arbitrary graph $G$ of size $N$, in such a way that the entry $A_{ij}$ of
$\textbf{A}$ is $1$ if $i$ and $j$ are adjacent in $G$, otherwise $A_{ij}$ is zero.  The degree $d_i$ of node $i$ is defined as $d_i \equiv \sum_{j=1}^N A_{ij}$.
Further, let $\mathbf{Z}$ be the diagonal matrix, whose entries are given by $Z_{ij} = d_i \delta_{ij}$ (here $\delta_{ij}$ is the Kronecker delta).
The Laplacian matrix is then defined as
\begin{equation} \label{eq:Laplacian}
\mathbf{L} \equiv  \mathbf{Z} - \mathbf{A}.
\end{equation}

Given that $\mathbf{L}$ is semi-definite positive, the Laplacian eigenvalues are all real and non-negative. Also, they are contained in the interval $[0, \min \{ N,  2 d_{\max} \} ]$, where $d_{\max} \equiv \max_i ( d_i)$. The set of all $N$ Laplacian eigenvalues $\lambda_1 \leq \lambda_{2} \leq ... \leq \lambda_N$ is called the Laplacian spectrum.
Following the definition (\ref{eq:Laplacian}), the smallest eigenvalue $\lambda_1$ is always null and corresponds to an eigenvector with all entries equal to $1$, according to the Perron-Frobenius theorem; the multiplicity of $0$ as an eigenvalue of $\mathbf{L}$ corresponds to the number of components of $G$.
The smallest non-zero Laplacian eigenvalue is often referred to as ``spectral gap'' and it provides information on the effective bipartitioning of a graph.
Basically, in a graph with a relatively small first non-trivial Laplacian eigenvalue the number of edges
that need to be cut away to generate a bipartition is relatively small; conversely, a large spectral gap characterizes
non-structured networks, with poor modular structure (see e.g., \cite{Girvan-PNAS2002,Donetti-JStat2004}).
The spectral gap is also associated
with spreading efficiency (random walks move around quickly and disseminate fluently
on graphs with large spectral gap) and with synchronizability (the stability of fully synchronized states on networks is enhanced when the spectral gap is large)  \cite{Donetti-JStat2004,Lovasz-1993}.

\begin{figure}
\begin{center}
\includegraphics[scale=0.27]{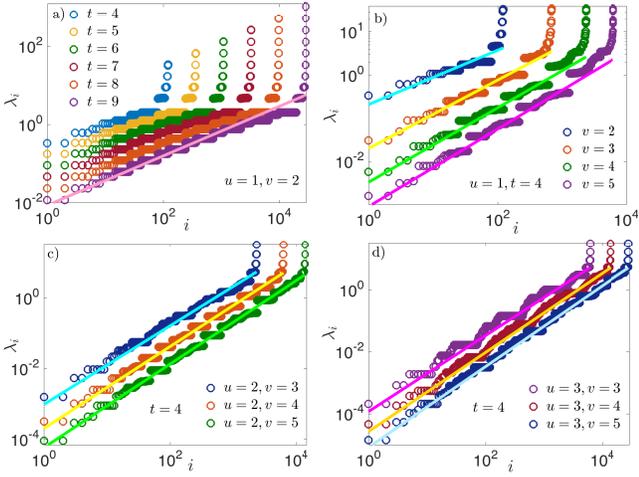}
\caption{In these panels we show in a logarithmic scale the Laplacian eigenvalues $\lambda_i$ sorted in ascending order, for different values of $u$ and $v$, shown in different colors as explained in the legend. The solid lines represent the best fit according to Eq.~(\ref{eq:lambda}). More precisely, in panel $a$ we focused on the case $u=1$ and $v=2$, while we explored different sizes: as expected, the slope of the best fit curve is independent of the network generation $t$; in panel $b$ we keep $u=1$ and $t=4$ fixed, while we let $v$ vary: as expected, the slope of the best fit curve grows with $v$; in panel $c$ we keep $u=2$ and $t=4$ fixed, while we let $v$ vary: as expected, the slope of the best fit curve grows with $v$; in panel $d$ we keep $u=3$ and $t=4$ fixed, while we let $v$ vary: as expected, the slope of the best fit curve grows with $v$. In any case, the best fit is performed by focusing on the small eigenvalues.}
\label{fig:laplacian}
\end{center}
\end{figure}

Let us now resume the $(u,v)$-flowers and look at their Laplacian spectrum. We derive numerically the spectrum for several choices of the parameters $u$ and $v$, and for several generations $t$. A convenient way to check their distribution is to sort eigenvalues in such a way that $\lambda_i \leq  \lambda_{i+1}$ and to look at the rate of growth of $\lambda_i$ as the index $i$ is increased. In fractal structures endowed with spectral dimension $d_s$ the scaling is given by (see e.g., \cite{Mieghen-2011})
\begin{equation} \label{eq:lambda}
\lambda_i  \sim i^{2/d_s}, ~~~ i=2,3,...
\end{equation}
at least as far as the lower part of the spectrum is concerned.
Our numerical results are collected in Fig.~\ref{fig:laplacian}.
As expected, $(u,v)$-flowers with $u>1$ confirm the previous scaling with $d_s = 2 \log(u+v)/\log(u v)$.
As for networks with $u=1$, we can still outline a power-law scaling with respect to the index $i$, although this time deviation in the upper part of the spectrum are more evident. By fitting the linear part, we obtain an exponent compatible with $\tilde{d}_s = 2 \log(1+v) / \log(v)$.
The characterization of the spectral density can therefore be accomplished in terms of the spectral dimension also for transfractal $(1,v)$-flowers.

 \begin{figure}
\begin{center}
\includegraphics[scale=0.27]{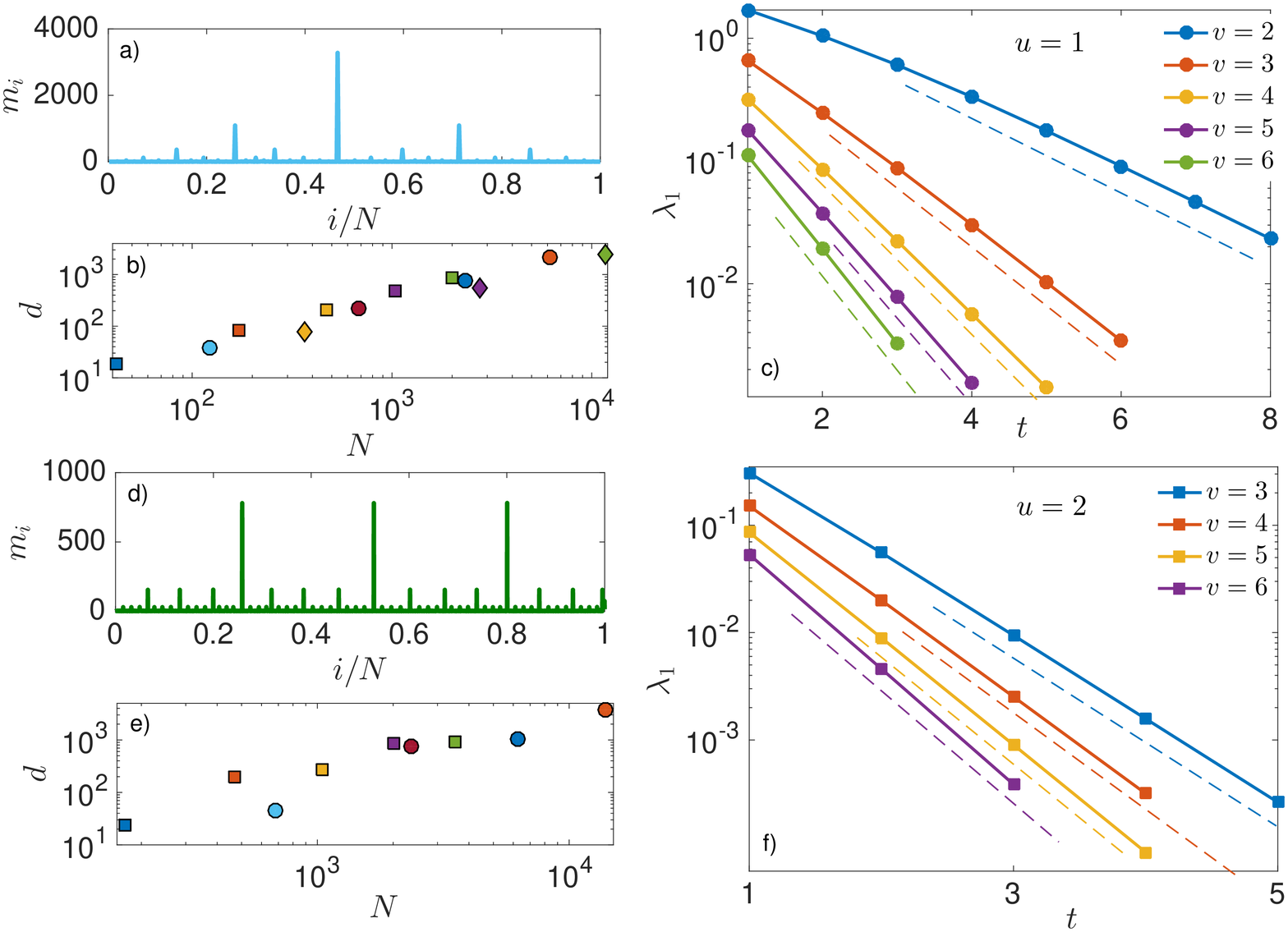}
\caption{
Panel $a$: The multiplicity $m_i$ of the eigenvalue $\lambda_i$ is plotted versus the normalized index $i/N$. Here we considered the case $u=1$, $v=2$, $t=8$. Notice that the structure of this plot mirrors the self-similarity of the network itself.
Panel $b$: The number $d$ of distinct eigenvalues is plotted versus the graph size $N$, for different choices of $v$ and $t$, while $u=1$ is kept fixed. In particular, we considered $t=3$ ($\square$) with $v \in [2,6]$, $t=4$ ($\bullet$) with $v \in [2,5]$, and $t=5$ ($\diamond$) with $v \in [2,4]$. Notice that, in general, $d$ depends on $u$, $v$, and $t$, and it grows algebraically with $N$.
Panel $c$: the spectral gap $\lambda_2$ for the $(1,v)$-flowers is plotted versus the generation $t$; several choices of $v$ are considered as explained by the legend. Notice the exponential decrease of the gap with $t$, which corresponds to an algebraic decrease with respect to the volume $N_t$.
Panels $d$-$f$ shows analogous figures for $u=2$.
In particular, in panel $d$ we considered the case $v=3$, $t=6$ and in panel $e$ we considered $t=3$ ($\square$) with $v \in [2,6]$, $t=4$ ($\bullet$) with $v \in [2,5]$. In panels $c$ and $f$ the dashed lines highlight the scaling $(uv)^{-t}$, as explained in the main text.}
\label{fig:laplacian2}
\end{center}
\end{figure}

In order to get further insight into the Laplacian spectrum, we plotted the eigenvalues multiplicity $m_i$ versus $i$ (see Fig.~\ref{fig:laplacian2}, panels $a$ and $d$), and the number $d$ of distinct eigenvalues versus $N$ (see Fig.~\ref{fig:laplacian2}, panels $b$ and $e$). For both cases $u=1$ and $u>1$ we see that $m_i$ mirrors the self-similarity of the network and $d$ grows approximately algebraically with the size.

Finally, we consider the spectral gap $\lambda_2$ and try to get insights into its dependence on the system parameters, by recalling that its inverse corresponds to the relaxation time $\tau$, which, in turn, is related to the mixing time, namely the characteristic time for reaching the stationary distribution (see e.g., \cite{Mieghen-2011,Donetti-JStat2004,Agl16}).
This can be estimated as the time necessary to cover a distance comparable with the system diameter $L_t$. For $u>1$, one has $L_t \sim u^t$ \cite{Bollt05,Dorogovtsev02}, in such a way that $\tau \sim L_t^{d_w}$. Therefore, we get $\lambda_2 \sim \tau^{-1} \sim u^{-t d_w}$ and, recalling $d_w = \log(uv)/\log(u)$, we get $\lambda_2 \sim (uv)^{-t} \sim 1/N$. As for the case $u=1$, one has $L_t \sim (v-1)t$ \cite{Bollt05,Dorogovtsev02} and, due to the transfractal nature of the graph, the characteristic time to cover a certain distance scales exponentially with the distance itself. Therefore, recalling $\tilde{d}_w = \log(v)/ (v-1)$, we get $\lambda_2 \sim (v)^{-t} \sim 1/N$. Remarkably, the explicit scalings obtained for $u>1$ and $u=1$ are analogous and this is checked numerically (see Fig.~\ref{fig:laplacian2}, panels $c$ and $f$).

\section{Conclusions}
\label{sec:4}
In this work we considered $(u,v)$-flowers which constitute a class of deterministic graphs: as the parameters $u$ and $v$ are tuned, the resulting graph exhibits different topological properties. In particular, when $u>1$ the graph is a fractal endowed with fractal dimension $d_f$, walk dimension $d_w$, and spectral dimension $d_s = 2 d_f /d_w = 2 \log(u+v)/\log(uv)$, and they are all finite. On the other hand, when $u=1$ the graph is transfractal, with transfractal dimension $\tilde{d}_f$,  transwalk dimension $\tilde{d}_w$, and spectral dimension $\tilde{d}_s = 2 \tilde{d}_f /\tilde{d}_w= 2 \log(1+v)/\log(v)$ (which coincides with $d_s$ when evaluated in $u=1$, without any singularity).
In fractals, the above mentioned dimensions are known to control the scaling of observables associated to dynamical processes. In this work we investigated the robustness of these scalings when moving from $u>1$ to $u$=1.
To this aim we considered first-passage problems for random walks embedded in $(u,v)$-flowers and the Laplacian spectrum of $(u,v)$-flowers. As expected, when $u>1$, we recover the well-known scalings; in particular, the first and the second moments of the global first-passage time in the limit of large size scale, respectively, as $\langle GFPT_t\rangle \sim N^{{2}/{d_s}}$ and  $\langle GFPT^2 \rangle \sim N^{{4}/{d_s}}$, also, the Laplacian eigenvalue density scales as $P(\lambda) \sim \lambda^{d_s/2 -1}$, in the limit $\lambda \rightarrow 0$.
Interestingly, when $u=1$ the same scalings are recovered by using $\tilde{d}_s$, as long as the target node has a rather large degree. This means that the average time to reach a hub scales analogously with the volume, independently on whether the exploration is compact or not.
Moreover, in the large size limit and regardless of whether $u$ is equal to or greater than $1$, we found that the spectral gap scales as the inverse of the system size $N$.

The investigations performed in this work can be further extended to account for the role of $\tilde{d}_s$ also for other phenomena, such as synchronizability \cite{Donetti-JStat2004} and number of spanning trees which are as well related to the distribution of eigenvalues in the Laplacian matrix. Moreover, towards realistic applications, it would be interesting to check the robustness of these results in the presence of some degree of noise, e.g., modeled in terms of stochastic weights on links or of stochastic deletion of a fraction of links.

\acknowledgments
{
Financial supports from China Scholarship Council (Grant No. 201708440148) and from GNFM-INdAM (``Progetto Giovani 2016'') are kindly acknowledged.
}

\appendix

 \section{Properties of probability generating functions}
\label{sec:PGF}
The probability generating function, defined by Eq.~(\ref{Def_PGF}), is  determined by the probability distribution and, in turn, it uniquely determines the probability distribution.
 If $T_1$ and $T_2$ are two random variables with the same probability generating function, they have the same probability distribution. Given the probability generating function $\Phi_{T}(z)$ of the random variable $T$, we can obtain the probability distribution $p_k$ ($k=0,1,2,\cdots$)  as the coefficient of $z^k$ in the Taylor's series expansion of $\Phi_{T}(z)$ about $z=0$. 

Also, when $\Phi_{T}(z)$ and its $n$-th order derivatives are well defined at $z = 1$, the $n$-th moment $\langle T^n \rangle \equiv \sum_{k=0}^{+\infty} k^n p_k$ can be written in terms of a combination of derivatives of $\Phi_{T}(z)$ evaluated at $z=1$\footnote{Notice that the radius of convergence of a probability generating function must be at least $1$, and, in particular, the normalizaton of $p_k$ yields to $\Phi_{T}(z)=1$. Also, a diverging derivative as $z \rightarrow 1$ means that the first moment is diverging as well, and similarly for higher order moments. For the quantities considered here (i.e., FPT, FRT, GFPT), given the finiteness of the underling structure, the moments are all finite. We also recall that he probability generating function of $p_k$ can also be see as the (discrete) Laplace transform of $p_k$. In this perspective the discrete function to be transformed does not need to be normalized.}. For example,
\begin{eqnarray}\label{n_moment}
\langle T \rangle  &=& \frac{d \Phi_{T}(z)}{ d z} \Big |_{z=1},\\
\langle T^2 \rangle  &=&  \frac{d^2 \Phi_{T}(z)}{ d z^2} \Big|_{z=1} + \frac{d \Phi_{T}(z)}{ d z} \Big|_{z=1}.
 \end{eqnarray}

Finally, we recall a couple of useful properties of the probability generating function (see e.g., \cite{Gut05}):\\
$i)$ Let $T_1$ and $T_2$ be two independent random variables with probability generating functions $\Phi_{T_1}(z)$ and $\Phi_{T_2}(z)$, respectively. Then, the probability generating function of the random variable $T_1+T_2$ reads as
  \begin{equation}\label{Sum_PGF2}
    \Phi_{T_1+T_2}(z)=\Phi_{T_1}(z)\Phi_{T_2}(z).
 \end{equation}
$ii)$ Let $N$, $T_1$, $T_2$, $\cdots$ be independent random variables. If $T_i$ ($i=1, 2, \cdots$) are identically distributed, each with probability generating function $\Phi_{T}(z)$, and $\Phi_{N}(z)$ is the probability generating function of $N$, then the random variable $S_N=T_1+T_2+\cdots+T_N$
has probability generating function
  \begin{equation}\label{Sum_PGFn}
    \Phi_{S_N}(z)=\Phi_{N}(\Phi_{T}(z)).
 \end{equation}

\section{Derivation of $\Phi_{FPT}(1,z)$ and related quantities}
\label{PGF_FPTg1}
The $(u,v)$-flower of generation $1$ is a ring with $w=u+v$ nodes, which are labeled as $H_0$, $H_1$, $H_2$, $\cdots$, $H_{w-1}$. Note that all the nodes of the network are equivalent to one another, in such a way that the FPT between two nodes depends only on their chemical distance, but not on the particular couple chosen. In particular, the FPT from $H_0$ to $H_u$ is the same as the FPT from $H_{w-u-1}$ to $H_{w-1}$.

In order to calculate  the probability generating function for the FPT from $H_{w-u-1}$ to $H_{w-1}$, we
assume that $H_{w-1}$ is a trap, that is, the transition probability from $H_{w-1}$ to $H_{k}$ be $0$ for any $k$.  Thus, the time taken by  a random walker, started at $H_{w-u-1}$, to reach $H_{w-1}$ is the FPT from $H_{w-u-1}$ to $H_{w-1}$. Let $P_k(n)$ be the  probability that the random walker is found at $H_k$ at time $n$ and let $\Phi_{k}(z)$ denote the generating function of $P_k(n)$ $(n=0,1,...)$. Since $H_{w-1}$ is a trap, when the walker reaches $H_{w-1}$ it is trapped and removed from the system and $P_{w-1}(n)$ corresponds to the probability that the walker first reaches $H_{w-1}$ at time $n$.
Therefore, $\Phi_{w-1}(z)$ is the probability generating function of the FPT from $H_{w-u-1}$ to $H_{w-1}$ (or, equivalently, from $H_0$ to $H_u$). \\
By definition, at time $n=0$,
\begin{equation}
\label{Pk0}
  P_{k}(0)=\left\{ \begin{array}{ll} 1 & k=w-u-1\\ 0 & k \neq w-u-1 \end{array} \right.,
\end{equation}
and, for $n>0$,
\begin{equation}
\label{Pkn}
  P_{k}(n)=\left\{ \begin{array}{ll} \frac{1}{2}P_{1}(n-1) & k=0 \\ \frac{1}{2}P_{w-3}(n-1) & k=w-2 \\ \frac{1}{2}[P_{0}(n-1)+P_{w-2}(n-1)] & k=w-1 \\ \frac{1}{2}[P_{k-1}(n-1)+P_{k+1}(n-1)] & \textrm{otherwise} \end{array} \right..
\end{equation}
Therefore,
\begin{equation}
\label{Phik}
  \Phi_{k}(z)=\left\{ \begin{array}{ll} \frac{z}{2}\Phi_{1}(z) & k=0 \\ \frac{z}{2}[\Phi_{k-1}(z)+\Phi_{k+1}(z)]+1 & k=w-u-1 \\ \frac{z}{2}\Phi_{w-3}(z) & k=w-2 \\ \frac{z}{2}[\Phi_{w-2}(z)+\Phi_{0}(z)] & k=w-1 \\ \frac{z}{2}[\Phi_{k-1}(z)+\Phi_{k+1}(z)] & \textrm{otherwise} \end{array} \right..
\end{equation}
Then, we obtain linear equations for the vector $\Phi=(\Phi_{0}(z), \Phi_{1}(z),...,\Phi_{w-1}(z))$, which, in matricial notation reads as
$\tilde{\Lambda} \cdot \Phi = b$, where $b$ is a vector whose entries are all null, except the $(w-u)$-th entry which is equal to $1$.
In order to solve the system it is convenient to introduce the so-called augmented matrix ${\Lambda}$ which is obtained by merging $\tilde{\Lambda}$ and $b$. More precisely, we obtain $\Lambda$ by inserting an additional column in $\tilde{\Lambda}$: this additional column appears in the $(w+1)$-th column of the augmented matrix, where every entry is $0$ except for the $(w-u)$-th entry which is $1$. This makes the last entry of the $(w-u)$-th row of the matrix $1$.
Thus, the augmented matrix $\Lambda=(\Lambda_{i,j})_{w\times(w+1)}$ is\\
$
\Lambda=
\left (
\begin{array}{ccccc|cccc|c}
        1            & -\frac{z}{2} & 0             & \cdots        & 0  & 0            & \cdots  & 0                 & 0 &0\cr
        -\frac{z}{2} & 1            & -\frac{z}{2}  & \cdots        & 0  & 0            & \cdots  & 0                 & 0 &0\cr
        0           & -\frac{z}{2} & 1              & \cdots        & 0  & 0            & \cdots  & 0                 & 0 &0\cr
                     &              &                &            &\ddots &              &         &                   &   & \cr
         0           & 0            & \cdots         & -\frac{z}{2}  & 1  & -\frac{z}{2} & \cdots  & 0                 & 0 &1\cr
                     &              &                &               &    &  \ddots      &         &                   &   & \cr
         0           & 0            & 0              & \cdots        & 0  & \cdots       & -\frac{z}{2} & 1            & 0 &0\cr
         -\frac{z}{2}& 0            & 0              & \cdots        & 0  & 0            & \cdots       & -\frac{z}{2} & 1 &0
\end{array}
\right ),$
where we highlighted by vertical lines the $(w-u)$-th column and the $w$-th column.
We now transform the $(i+1)$-th row of the matrix $\Lambda$ according to $\Lambda_{i+1,j} \rightarrow \Lambda_{i+1,j} + (z/2) \Lambda_{i,i} \Lambda_{i,j}$, for $j=1, ..., w+1$ and $i=1,...,w-1$.  That is, we replace the $(i+1)$-th row with the (properly weighted) sum of the rows $i$ and $i+1$. As a consequence, $\Lambda_{i+1,i}=0$. We note that, after such operations, the  diagonal elements of $\Lambda$ are transformed into
 $\Lambda_{1,1}=1$, $\Lambda_{2,2}=1-\frac{z^2}{4}$, $\Lambda_{w,w}=1$, and $\Lambda_{i,i}=\frac{\delta_i}{\delta_{i-1}}$, for $1<i<w$,
 where ${\delta_i}$ satisfies  the recurrence relation 
\begin{equation}
\label{Rec_delta_i}
  \delta_i=\delta_{i-1}-\frac{z^2}{4}\delta_{i-2},
\end{equation}
with initial conditions $\delta_1=1$, $\delta_2=1-\frac{z^2}{4}$.
We also find that
\begin{equation}
\label{LambdaW}
  \Lambda_{i,w+1}=\left\{ \begin{array}{ll} 0 & i<w-u \\
                                         1 & i=w-u \\
                                          \left (\frac{z}{2} \right)^{i-w+u}\frac{\delta_{w-u-1}}{\delta_{i-1}} & w-u<i\leq w \end{array} \right..
\end{equation}
We now further transform the augmented matrix $\Lambda$ by summing $-\frac{\Lambda_{w,i}}{\Lambda_{i,i}} \Lambda_{i,j}$ to each element $\Lambda_{w,j}$ of the $w$-th row. We repeat the same operation sequentially from $i=1$ to $i=w-1$, to obtain an upper triangular matrix with  $\Lambda_{w,w}=1$. The probability generating function $ \Phi_{FPT}(1,z)$ that we are looking for corresponds to $\Phi_{w-1}(z)$, which, in turn, is provided by the entry $(w,w+1)$ in the upper triangular matrix just found. Therefore, we can write
\begin{equation}
  \Phi_{FPT}(1,z)= \left(\frac{z}{2}\right)^u\frac{\delta_{w-u-1}}{\delta_{w-1}}+\sum_{i=w-u}^{w-1} \left(\frac{z}{2} \right)^{2i-w+u}\frac{\delta_{w-u-1}}{\delta_{i}\delta_{i-1}}.
\end{equation}
In order to obtain an exact expression for $\Phi_{FPT}(1,z)$, we need to solve the discrete difference equation in Eq.~(\ref{Rec_delta_i}). Its characteristic equation reads as $\lambda^2-\lambda+\frac{z^2}{4}=0$
with two different characteristic roots $\lambda_{1,2}=\frac{1\pm\sqrt{1-z^2}}{2}$. Therefore~\cite{Richard_Bru01}
\begin{equation}
\label{delta_i}
  \delta_i=c_1\times \left(\frac{1+\sqrt{1-z^2}}{2} \right)^i+c_2\times \left (\frac{1-\sqrt{1-z^2}}{2} \right)^i,
\end{equation}
where $c_1=\frac{1-z^2+\sqrt{1-z^2}}{2(1-z^2)}$ and $c_2=\frac{1-z^2-\sqrt{1-z^2}}{2(1-z^2)}$, which are determined by  the initial conditions.

In Appendix C, the following derivatives of $\Phi_{FPT}(1,z)$ evaluated at $z = 1$ are used:
\begin{equation}
\label{M1FPT}
 \langle FPT_1\rangle=\left.\frac{d}{d z} \Phi_{FPT}(1,z)\right|_{z=1}=uv ,
\end{equation}
\begin{equation}
\label{D2FPT}
 \left.\frac{d^2}{d z^2} \Phi_{FPT}(1,z)\right|_{z=1}=\frac{uv}{3}(u^2+3uv+v^2-5).
\end{equation}

\section{Derivation of Eqs.~\ref{Rec_PGF_FPTm}-\ref{M2FPTm}}
 \label{App:PGF_FPTt}

First, we derive Eq.~(\ref{Rec_PGF_FPTm}), which depicts the recurrence relation for the probability generating function of the FPT from $H_0$ to $H_u$.
To this aim it is convenient to imagine that there is a trap located at hub $H_u$.
We denote with $\pi$ any first-passage path from $H_0$ to $H_u$ on $G(t)$ ($t>1$); its length is $F_{H_0\rightarrow H_u}(t)$, which is the FPT from $H_0$ to $H_u$. Let  $v_i$ be a node within $G(t)$ reached at time $i$. Then the path can be rewritten as $\pi=(v_0=H_0,v_1,v_2\cdots,T_{H_0\rightarrow H_u}(t)=H_u).$
We denote with $\Omega$ the set of all the hub nodes $\{H_0, H_1, H_2, \cdots, H_{w-1} \}$  and introduce the observable $\tau_i=\tau_i(\pi)$,
which is defined recursively as follows:
\begin{eqnarray} \label{OB}
\left\{
  \begin{array}{ll}
   \tau_0(\pi)=0,\\
   \tau_i(\pi)=\min\{k: k>\tau_{i-1}, v_{k} \in \Omega, v_{k} \neq v_{\tau_{i-1}} \}.
   \end{array}
   \right.
\end{eqnarray}
Then,  considering only nodes in the set $\Omega$, the path $\pi$ can be restated into a ``simplified path'' defined as $\iota(\pi)=(v_{\tau_{0}}=H_0,v_{\tau_1},\cdots, v_{\tau_L}=H_u )$,
where $L$ is the total number of times that, along the path $\pi$, we move from one hub to another. In fact, the ``simplified path'' is obtained by removing any nodes of $\pi$ except the hub nodes. 
Note that $\Omega$ represents  the node set of the $(u,v)$-flower with generation $1$ and the path $\iota(\pi)$ includes only the  nodes of $\Omega$. Thus, $\iota(\pi)$ is a first-passage path from $H_0$ to $H_u$  on the  $(u,v)$-flower with generation $1$ and $L$ is the FPT from $H_0$ to $H_u$  on the $(u,v)$-flower with generation $1$. Therefore, the $L$ probability generating function is $ \Phi_{FPT}(1,z)$.

Let  $T_i$ ($i=1,2,\cdots, L$) be the time taken to move from $v_{\tau_{i-1}}$ to $v_{\tau_{i}}$, namely $T_i=\tau_i-\tau_{i-1}$. Therefore, 
\begin{equation}
\label{FPT_pathlength}
  F_{H_0\rightarrow H_u}(t)= T_1+T_2+\cdots+T_L.
\end{equation}

As shown in Fig.~\ref{Self_similar}, $G(t)$ is composed of $w$ copies of $G(t-1)$ which are labeled as  $\Gamma_1$,  $\Gamma_2$,  $\cdots$, $\Gamma_w$. The
$w$ hub nodes of $G(t)$ can also be regarded as the two special hub nodes (i.e., $H_0$ and $H_u$) of $G(t-1)$. For example, $H_0$ and $H_1$ can be regarded as $H_0$ and $H_u$ of $\Gamma_1$; $H_1$ and $H_2$ can be regarded as $H_0$ and $H_u$ of $\Gamma_2$. Therefore, for any $i=1, 2,\cdots$, $L$, $v_{\tau_{i-1}}$ and $v_{\tau_i}$ can be regarded as the two special hub nodes (i.e., $H_0$ and $H_u$) of $G(t-1)$ and  $T_i$, which is the FPT from $v_{\tau_{i-1}}$ to $v_{\tau_i}$, can be regarded as the FPT from  $H_0$ to $H_u$ on $G(t-1)$.  Then, the $T_i$ probability generating function is $ \Phi_{FPT}(t-1,z)$.

Note that  $L$,  $T_1$, $T_2$, $\cdots$  are independent random variables. According to the properties of the  probability generating function presented in Appendix~\ref{sec:PGF} (see Eqs.~(\ref{Sum_PGF2}) and (\ref{Sum_PGFn})), we have, for any $t>1$,
\begin{equation}\label{Rec_PGF_FPT}
  \Phi_{FPT}(t,z)=\Phi_{FPT}(1,\Phi_{FPT}(t-1,z)).
\end{equation}
Therefore, Eq.~(\ref{Rec_PGF_FPTm}) is obtained. Let us now focus on the derivation of Eqs.~(\ref{MFPT_Em}) and (\ref{M2FPTm}).

For $t=1$, Eq.~(\ref{M1FPT}) shows that Eq.~(\ref{MFPT_Em}) holds. 

For any $t>1$, by taking the first order derivative on both sides of Eq.~(\ref{Rec_PGF_FPT}) and posing $z=1$,
we obtain 
\begin{eqnarray}
    \langle FPT_t \rangle
                 &=& \left. \frac{d}{d z}\Phi_{FPT}(1,z)\right|_{z=1}\times\left. \frac{\partial}{\partial z}\Phi_{FPT}(t-1,z)\right|_{z=1}\nonumber \\
                 &=&\langle FPT_1 \rangle \langle FPT_{t-1} \rangle               \nonumber \\
                 \label{MFPT}
                 &=&\cdots=\langle FPT_1 \rangle^{t}=(uv)^{t}.
\end{eqnarray}
Therefore Eq.~(\ref{MFPT_Em}) holds for any $t \geq 1$.

We can similarly derive Eq.~(\ref{M2FPTm}). 
 Let $\Theta_t\equiv\left. \frac{\partial^2}{\partial z^2}\Phi_{FPT}(t,z)\right|_{z=1}$.
By taking the second order derivative on both sides of Eq.~(\ref{Rec_PGF_FPT}) and posing $z=1$, for any $t\geq1$,
\begin{eqnarray}\label{Rec_d2FPT}
   \Theta_t
                 &=& uv\Theta_{t-1}+\Theta_{1}(uv)^{2t-2} = \cdots             \nonumber \\
                 &=&(uv)^{t-1}\Theta_{1}+\Theta_{1}\left[(uv)^{t}+(uv)^{t+1}+\cdots+(uv)^{2t-2}\right]\nonumber \\
                 &=&(u^2+v^2+3uv-5)\frac{(uv)^{t}-1}{3(uv-1)}.
\end{eqnarray}
Therefore, for any $t\geq1$, 
\begin{eqnarray}
   \langle FPT^2_t\rangle&=&\Theta_t+\langle FPT_t \rangle\nonumber \\
                 &=&(u^2+v^2+3uv-5)\frac{(uv)^{t}-1}{3(uv-1)}+(uv)^{t}.\nonumber
\end{eqnarray}
Therefore Eq.~(\ref{M2FPTm}) is obtained for $t\geq 1$. As for the case $t=0$, recalling that $G(0)$ is made by two nodes connected by an edge, we have $\langle FPT^2_0\rangle=1$. Therefore, Eq.~(\ref{M2FPTm})  also holds for $t=0$.

\section{Recursive equations for the return time probability generating function for a given hub} \label{App:RT}
Given that all the main hubs in $G(t)$ are topologically equivalent, without loss of generality, we can consider the return time for the hub $H_0$. Let $T_{H_0\rightarrow H_0}(t)$ denote the return time to $H_0$ in $G(t)$ (notice that this may not be the \emph{first} time that the random walker returns to $H_0$) and  $\Phi_{RT}(t,z)$  denote its  probability generating function\footnote{Notice that the probability distribution for $T_{H_0\rightarrow H_0}(t)$ is not normalized over time and, accordingly, $\Phi_{RT}(t,z)$ is diverging for $z \rightarrow 1$. Nonetheless, one can exploit $\Phi_{RT}(t,z)$ to obtain closed-form, well-defined expression for $\Phi_{FPT}(t,z)$, $\Phi_{FRT}(t,z)$ and $\Phi_{GFPT}(t,z)$. See e.g., \cite{weiss,Agl16} for further details.}. For any $T_{H_0\rightarrow H_0}(t)$ $(t>0)$,
  we can construct independent random variables $T^a_{H_0\rightarrow H_0}(t)$, $L$,  $T_1$, $T_2$, $\cdots$, such that
\begin{equation}
\label{pathlengthd}
  T_{H_0\rightarrow H_0}(t)= T_1+T_2+\cdots+T_{L}+T^a_{H_0\rightarrow H_0}(t),
\end{equation}
where $T^a_{H_0\rightarrow H_0}(t)$ denotes the return time to $H_0$ in the presence of an absorbing hub $H_u$, $L$ is the return time to  $H_0$  on the network with generation $0$ and $T_i$ ($i=1,2,\cdots$, $L$) are identically distributed random variables, each of which can be regarded as the first-passage time from  hub $H_0$ to  hub $H_u$ on $G(t)$. Therefore, the $L$ probability generating function is $ \Phi_{RT}(0,z)$ and the $T_i$ ($i=1,2,\cdots$, $L$) probability generating function is  $ \Phi_{FPT}(t,z)$. If we denote the $T^a_{H_0\rightarrow H_0}(t)$ probability generating function by $ \Phi^a_{RT}(t,z)$, 
the return time $T_{H_0\rightarrow H_0}(t)$ probability generating function satisfies~\cite{Gut05}
\begin{equation}\label{RR_PGF_RTOd}
  \Phi_{RT}(t,z)=\left. \Phi_{RT}(0,x)\right|_{x=\Phi_{FPT}(t,z)}\times\Phi^a_{RT}(t,z).
\end{equation}
Similarily to the derivation of Eq.~(\ref{RR_PGF_RTOd}) , we find that, for any $t>1$,
 the $T^a_{H_0\rightarrow H_0}(t)$ probability generating function satisfies
\begin{equation}\label{Rec_PGF_RT_a}  
  \Phi^a_{RT}(t,z)=\Phi^a_{RT}(1,\Phi_{FPT}(t-1,z))\times\Phi^a_{RT}(t-1,z).
\end{equation}
Replacing $\Phi^a_{RT}(t,z)$ and $\Phi_{FPT}(t,z)$ from, respectively,  Eqs.~ (\ref{Rec_PGF_RT_a}) and (\ref{Rec_PGF_FPT}) in Eq.~ (\ref{RR_PGF_RTOd}),  we obtain the recurrence relation of $\Phi_{RT}(t,z)$, which can be simplified as
\begin{equation}\label{R_RTd}
  \Phi_{RT}(t,z)=\Phi_{RT}(t-1,z)/\psi(\Phi_{FPT}(t-1,z)),
\end{equation}
for any $t \geq 1$, with $\Phi_{RT}(0,z)=\frac{1}{1-z^2}$ and
\begin{equation}\label{Nrat}
\psi(z)\equiv\frac{\Phi_{RT}(0,z)}{\Phi_{RT}(1,z)}.
\end{equation}
 The derivation of Eqs.~(\ref{pathlengthd})-(\ref{R_RTd}) and $\psi(z)$
 are  presented in the following subsections.


\subsection{Derivation of Eqs.~(\ref{pathlengthd}), (\ref{RR_PGF_RTOd}) and (\ref{R_RTd})} 
 \label{sec:Rec_relation}

Let us consider an arbitrary return path $\pi$  starting from $H_0$ and ending at $H_0$ on $G(t)$. This can be written as $\pi=(v_0=H_0,v_1,v_2\cdots,v_{T_{H_0\rightarrow H_0}(t)}=H_0),$
 where $v_i$ is the node  reached at time $i$ and $T_{H_0\rightarrow H_0}(t)$ is the  length of the path $\pi$.
Let $\Omega$ denote  the set of nodes $\{H_0,H_u \}$ and introduce the observable $\tau_i=\tau_i(\pi)$,
which is defined recursively by Eq.~(\ref{OB}). 
Then,  considering only nodes in the set $\Omega$, the path $\pi$ can be restated into a ``simplified path'' defined as $\iota(\pi)=(v_{\tau_{0}}=H_0,v_{\tau_1},\cdots, v_{\tau_L}=H_0 )$,
where  $L=\max\{i: v_{\tau_i}=H_0\}$, which is the total number of observables obtained from the path $\pi$. 

Note that $\Omega$ represents  the node set of the $(u,v)$-flower with generation $0$ and the path $\iota(\pi)$ includes only the  nodes of $\Omega$. Thus, $\iota(\pi)$ is a return path of $H_0$  on the  $(u,v)$-flower with generation $0$ and $L$ is the return time of  $H_0$  on the $(u,v)$-flower with generation $0$. Therefore, the $L$ probability generating function is $ \Phi_{RT}(0,z)$.

For any return path  $\pi$ of $H_0$,   maybe $v_{\tau_L}$ is not the last node of path  $\pi$. That is to say, after node $v_{\tau_L}$, path  $\pi$ may include a sub-path from $H_0$ to $H_0$, which does not reach  $H_u$. In principle, the sub-path may include any node of $G(t)$ except $H_u$. Therefore, the sub-path can be regarded as a return path of $H_0$ in the presence of an absorbing hub $H_u$.  We denote its length by $T^a_{H_0\rightarrow H_0}(t)$  and denote its probability generating function by $ \Phi^a_{RT}(t,z)$.

Let $T_i$ ($i=1,2,\cdots, L$) be the time taken to move from $v_{\tau_{i-1}}$ to $v_{\tau_{i}}$, namely $T_i=\tau_i-\tau_{i-1}$. Therefore the return time $T_{H_0\rightarrow H_0}(t)$ on $G(t)$ satisfies
\begin{eqnarray}
\label{pathlength}
 T_{H_0\rightarrow H_0}(t) &=& \tau_L - \tau_0+T^a_{H_0\rightarrow H_0}(t) \nonumber \\
   &=& T_1+T_2+\cdots+T_L+T^a_{H_0\rightarrow H_0}(t),
\end{eqnarray}
and Eq.~(\ref{pathlengthd}) is obtained.

Because $v_{\tau_i}=H_0$ (or $H_u$) for any $i=0, 1,2,\cdots$, $L$ and $v_{\tau_i} \neq v_{\tau_{i-1}}$ for any $i>0$. Then $T_i$ ($i=1,2,\cdots$, $L$) are identically distributed random variables, each of them is the first-passage time from  hub $H_0$ to  hub $H_u$ (or from hub $H_u$ to  hub $H_0$) on $G(t)$. Its  probability generating function is  $ \Phi_{FPT}(t,z)$.

Note that  $L$, $T^a_{H_0\rightarrow H_0}(t)$, $T_1$, $T_2$, $\cdots$  are independent random variables. According to the properties  of the  probability generating function (see Eqs.~(\ref{Sum_PGF2}) and (\ref{Sum_PGFn})), the return time $T_{H_0\rightarrow H_0}(t)$ probability generating function satisfies
\begin{equation}\label{RR_PGF_RTO}
  \Phi_{RT}(t,z)=\left. \Phi_{RT}(0,x)\right|_{x=\Phi_{FPT}(t,z)}\times\Phi^a_{RT}(t,z).
\end{equation}
Therefore Eq.~(\ref{RR_PGF_RTOd}) is obtained.

Replacing $\Phi^a_{RT}(t,z)$ from Eq.~(D3) in Eq.~(\ref{RR_PGF_RTO}),
for any $t>0$,
 \begin{eqnarray}\label{DR_PGF_RT}
& &\Phi_{RT}(t,z) \nonumber \\
 &=&\Phi_{RT}(0,\Phi_{FPT}(t,z))\times \Phi^a_{RT}(1,\Phi_{FPT}(t\!-\!1,z))  \nonumber \\
     & & \times\Phi^a_{RT}(t\!-\!1,z)        \nonumber \\
     &=& \frac{\Phi_{RT}(0,\Phi_{FPT}(t,z))\times \Phi^a_{RT}(1,\Phi_{FPT}(t\!-\!1,z))}{\Phi_{RT}(0,\Phi_{FPT}(t\!-\!1,z))} \nonumber \\
     & & \times\Phi_{RT}(0,\Phi_{FPT}(t\!-\!1,z))\times \Phi^a_{RT}(t\!-\!1,z)\nonumber \\
     &=&\frac{\Phi_{RT}(0,\Phi_{FPT}(t,z))\times \Phi^a_{RT}(1,\Phi_{FPT}(t\!-\!1,z))}{\Phi_{RT}(0,\Phi_{FPT}(t\!-\!1,z))} \nonumber \\
     & & \times\Phi_{RT}(t\!-\!1,z).
\end{eqnarray}
Recalling Eq.~(\ref{Rec_PGF_FPT}), 
we can obtain $\Phi_{RT}(0,\Phi_{FPT}(t,z)) = \left. \Phi_{RT}(0,\Phi_{FPT}(1,x))\right|_{x=\Phi_{FPT}(t-1,z)}$.
 Thus,
 \begin{eqnarray}
 &&\Phi_{RT}(0,\Phi_{FPT}(t,z)) \times \Phi^a_{RT}(1,\Phi_{FPT}(t\!-\!1,z))  \nonumber \\
     &=&\left. \left[\Phi_{RT}(0,\Phi_{FPT}(1,x)) \times \Phi^a_{RT}(1,x)\right]\right|_{x=\Phi_{FPT}(t-1,z)} \nonumber \\
     &=& \left. \Phi_{RT}(1,x)\right|_{x=\Phi_{FPT}(t-1,z)}.
 \end{eqnarray}
Now, recalling from Eq.~(\ref{Nrat}) that $\psi(z)\equiv \Phi_{RT}(0,z)/ \Phi_{RT}(1,z)$, we have
 \begin{eqnarray} \label{deng}
&&\frac{\Phi_{RT}(0,\Phi_{FPT}(t,z))\times \Phi^a_{RT}(1,\Phi_{FPT}(t\!-\!1,z))}{\Phi_{RT}(0,\Phi_{FPT}(t\!-\!1,z))} \nonumber \\
&\equiv&\frac{1}{\left. \psi(x)\right|_{x=\Phi_{FPT}(t-1,z)}}.
\end{eqnarray}
Inserting  Eq.~(\ref{deng}) into Eq.~(\ref{DR_PGF_RT}), for any $t>0$, we obtain
\begin{equation}\label{R_RT}
  \Phi_{RT}(t,z)=\Phi_{RT}(t-1,z)/\psi(\Phi_{FPT}(t-1,z)),
\end{equation}
and Eq.~(\ref{R_RTd}) is obtained.

In the network of generation $0$, the return probability in odd times is $0$ and the return probability in even times is $1$, therefore $ \Phi_{RT}(0,z)=\sum_{n=0}^{+\infty}z^{2n}=1/(1-z^2)$.


\subsection{Derivation of Eq.~(\ref{Rec_PGF_RT_a})}
 \label{PGF_aRTt}
Let us consider an arbitrary return path $\pi$ to $H_0$ on $G(t)$ in the presence of an absorbing hub $H_u$. This path can be written as $\pi=(v_0=H_0,v_1,v_2\cdots,v_{T^a_{H_0\rightarrow H_0}(t)}=H_0),$
 where $v_i$ is the node  reached at time $i$ and $T^a_{H_0\rightarrow H_0}(t)$ is the length of the path $\pi$.
Let  $\Omega\equiv\{H_0,H_1,\cdots, H_{w-1}\}$. Similar to what was done in the previous Subsec.~\ref{sec:Rec_relation}, we introduce the observable $\tau_i=\tau_i(\pi)$,
which is defined recursively as in Eqs.~(\ref{OB}).
Then,  the path $\pi$ can be restated as a ``simplified path'' defined as $\iota(\pi)=(v_{\tau_{0}}=H_0,v_{\tau_1},\cdots, v_{\tau_{L}}=H_0)$
where  $v_{\tau_{i}}\in \Omega$ $(i=1,2,..., L)$, and $L$ is the total number of observables obtained from the path $\pi$.

For any return path  $\pi$ of $H_0$ in the presence of an absorbing hub $H_u$, similar to the discussion in Subsec.~\ref{sec:Rec_relation},  the path  $\pi$ includes a sub-path from $H_0$ to $H_0$ after node $v_{\tau_{L}}$. The sub-path does not reach any  node in  $\Omega$ except for $H_0$.  Therefore, the sub-path can be regarded as a return path of $H_0$ on $G(t)$ in the presence of $w-1$ absorbing hubs (i.e., $H_1$,  $H_2$, $\cdots$,   $H_{w-1}$). In fact, the sub-path only includes nodes of  $\Gamma_1$ and $\Gamma_w$ (see Fig.~4 of the main text), which are  copies of $G(t-1)$. By symmetry,  nodes of $\Gamma_1$ are in one to one correspondence with nodes of $\Gamma_w$. If we replace all the nodes of  $\Gamma_w$ with the corresponding nodes of $\Gamma_1$ in the sub-path, we obtain a return path of $H_0$ in $\Gamma_1$ which never reaches hub $H_1$. It is a return path of $H_0$ on $\Gamma_1$ in the presence of an absorbing hub  $H_1$  and has the same path length as the original sub-path. If we look at $\Gamma_1$ as a copy of $G(t-1)$, the hub $H_1$ of $G(t)$ can also be looked at as $H_u$ of $G(t-1)$.   Therefore, the  length of the sub-path  after node $v_{\tau_{L}}$ can be regarded as the return time of $H_0$ on $G(t-1)$ in the presence of an absorbing hub  $H_u$.

Let  $T_i=\tau_i-\tau_{i-1}$ for any $i=1,2,\cdots, {L}$. Therefore the length of the path $\pi$ satisfies
\begin{equation}
\label{RTa_pathlength}
  T^a_{H_0\rightarrow H_0}(t)= T_1+T_2+\cdots+T_L+ T^a_{H_0\rightarrow H_0}(t-1).
\end{equation}

Note that $\Omega$ represents  the node set of the $(u,v)$-flower with generation $1$ and the path $\iota(\pi)$ includes only the  nodes of $\Omega$. Thus, $\iota(\pi)$ is a return path of $H_0$  on $G(1)$ in the presence of an absorbing hub $H_u$ and $L$ is the path length. Therefore, the $L$ probability generating function is $ \Phi^a_{RT}(1,z)$.

Similarly, we find that $T_i$ ($i=1,2,\cdots$) are identically distributed random variables, each of them can be regarded as the first-passage time from  hub $H_0$ to $H_u$  on $G(t-1)$. Therefore  the $T_i$ ($i=1,2,\cdots$) probability generating function is $\Phi_{FPT}(t-1,z)$. According to the properties of the  probability generating function (see Eqs.~(\ref{Sum_PGF2}) and (\ref{Sum_PGFn})),  we  obtain the required Eq.~(\ref{Rec_PGF_RT_a}). 

\subsection{Derivation of $\psi(z)$ and related quantities}
\label{PGF_RT_g1}
We first derive $\Phi_{RT}(1,z)$, which is the return time probability generating function for a hub on the $(u,v)$-flower of generation $1$, and then $\psi(z)$ is obtained.

Note that the structure of a $(u,v)$-flower of generation $1$ is a ring with $w=u+v$ nodes, which are labeled as $H_0$, $H_1$, $H_2$, $\cdots$, $H_{w-1}$, and all the $w$ nodes are topologically equivalent. Thus, without loss of generality, $\Phi_{RT}(1,z)$ can be considered as the return time probability generating function for the hub $H_{w-1}$. Consider a random walk on the $(u,v)$-flower of generation $1$ devoid of traps and let $P_k(n)$ denote the  probability that the random walker is found at $H_k$ at time $n$ and $\Phi_{k}(z)$ denotes the generating function of $P_k(n)$ $(n=0,1,...)$. As a consequence, $\Phi_{w-1}(z)$ corresponds to $\Phi_{RT}(1,z)$. For $n=0$, one has $P_{k}(0)= \delta_{k,w-1}$, while for $n>0$,
\begin{equation}
\label{PknRT}
  P_{k}(n)=\frac{1}{2}[P_{k-1}(n-1)+P_{k+1}(n-1)],
\end{equation}
where $k=0,1,...,w-1$ and $k-1$,  $k+1$ are meant modulo $w$. 
Therefore
\begin{equation}
\label{PhikRT}
  \Phi_{k}(z)=\left\{ \begin{array}{ll}  \frac{z}{2}[\Phi_{w-2}(z)+\Phi_{0}(z)]+1 & k=w-1 \\
   \frac{z}{2}[\Phi_{k-1}(z)+\Phi_{k+1}(z)] & k \neq w-1 \end{array} \right..
\end{equation}
%
%
%
As anticipated, the $(u,v)$-flower of generation $1$ is a ring with $w=u+v$ nodes; in this embedding space, the probability to find the walker on a given site only depends on its distance from the starting node and, in particular, taking the random walker starting from $H_{w-1}$, we have $\Phi_{0}(z)\equiv \Phi_{w-2}(z)$, $\Phi_{1}(z)\equiv \Phi_{w-3}(z)$, $...$.

If $w$ is even, we have $\frac{w}{2}+1$ independent probability generating functions $\Phi_{k}(z)$ with $k=\frac{w}{2},\frac{w}{2}+1,...,w-1$, and Eq.~(\ref{PhikRT}) can be rewritten as the linear equation $\tilde{\Delta} \Psi=c$ for the vector $\Phi=(\Phi_{\frac{w}{2}}(z), \Phi_{\frac{w}{2}+1}(z),...,\Phi_{w-1}(z))$, where the vector $c$ has all null entries, except the last entry which is equal to $1$. Again, as in  Appendix~\ref{PGF_FPTg1}, we introduce the augmented matrix $\Delta=(\Delta_{i,j})_{(\frac{w}{2}+1)\times(\frac{w}{2}+2)}$, of the form
 \begin{equation}
 \Delta=
 \left(\begin{array}{cccccccc}
          1            & -{z}         & 0             & \cdots             & 0     & 0                 & 0 &0\\
          -\frac{z}{2} & 1            & -\frac{z}{2}  & \cdots             & 0     & 0                 & 0 &0\\
          0           & -\frac{z}{2} & 1              & \cdots             & 0     & 0                 & 0 &0\\
                      &              &                & \ddots            &       &                   &   & \\
          0           & 0            & 0              & \cdots             & -\frac{z}{2} & 1          &-\frac{z}{2} &0\\
          0           & 0            & 0              & \cdots             & 0     & -z                & 1 &1
\end{array} \right). \nonumber
\end{equation}
Similarly to the procedure applied to $\Lambda$ in Appendix \ref{PGF_FPTg1}, we transform the $(i+1)$-th row of $\Delta$ according to $\Delta_{i+1,j} \rightarrow \Delta_{i+1,j} + \frac{z}{2} \Delta_{i,i} \Delta_{i,j}$, for $j=1,2,...,\frac{w}{2}+1$ and $i=1, 2,...,\frac{w}{2}$. These operations transform the augmented matrix $\Delta$ to an upper triangular matrix,  with
 $\Delta_{1,1}=1$, $\Delta_{2,2}=1-\frac{z^2}{2}$, $\Delta_{i,i}= \alpha_i / \alpha_{i-1}$ for $1< i \leq \frac{w}{2}$, and $\Delta_{\frac{w}{2}+1,\frac{w}{2}+1}= 1 - \frac{z^2}{2} \alpha_{\frac{w}{2}-1} / \alpha_{\frac{w}{2}}$,
   where ${\alpha_i}$ satisfies  the same recurrence relation as shown in Eq.~(\ref{Rec_delta_i}), with the alternate initial conditions $\alpha_1=1$ and $\alpha_2=1-\frac{z^2}{2}$. Therefore $\alpha_i$ has the same general formula as Eq.~(\ref{delta_i})~\cite{Richard_Bru01} and we can further obtain $c_1=c_2=1$ by using the initial conditions. Thus 
\begin{equation}
\label{alpha_i}
  \alpha_i= \left (\frac{1+\sqrt{1-z^2}}{2} \right)^i+ \left(\frac{1-\sqrt{1-z^2}}{2}\right)^i.
\end{equation}
Note that $\Delta_{\frac{w}{2}+1,\frac{w}{2}+2}=1$. Therefore, if $w$ is even,
  \begin{equation}
 \label{PGF_RT_E}
 \Phi_{RT}(1,z)=\frac{\Delta_{\frac{w}{2}+1,\frac{w}{2}+2}}{\Delta_{\frac{w}{2}+1,\frac{w}{2}+1}}=\frac{\alpha_{\frac{w}{2}}}{\alpha_{\frac{w}{2}}-\frac{z^2}{2}\alpha_{\frac{w}{2}-1}}.
 \end{equation}

If $w$ is odd, we  have $\frac{w+1}{2}$ independent probability generating function $\Phi_{k}(z)$ with $k=\frac{w-1}{2},\frac{w+1}{2},...,w-1$, and Eq.~(\ref{PhikRT}) can be rewriten as the linear equations of $\Phi=(\Phi_{\frac{w-1}{2}}(z), \Phi_{\frac{w+1}{2}}(z),...,\Phi_{w-1}(z))$, with
 augmented matrix $\Pi=(\Pi_{i,j})_{(\frac{w+1}{2})\times(\frac{w+3}{2})}$, expressed as
 \begin{equation}
 \Pi=
 \left(\begin{array}{cccccccc}
       1-\frac{z}{2}  & -\frac{z}{2} & 0             & \cdots             & 0     & 0                 & 0 &0\\
          -\frac{z}{2} & 1            & -\frac{z}{2}  & \cdots             & 0     & 0                 & 0 &0\\
          0           & -\frac{z}{2} & 1              & \cdots             & 0     & 0                 & 0 &0\\
                      &              &                & \ddots            &       &                   &   & \\
          0           & 0            & 0              & \cdots         &-\frac{z}{2} & 1         & -\frac{z}{2} &0\\
          0           & 0            & 0              & \cdots             & 0      & -z              & 1   &1
\end{array} \right). \nonumber
\end{equation}
\newline
Again, we handle the augmented matrix by replacing the $(i+1)$-th row with the sum of the $i$-th row times $\frac{z}{2}\Pi_{i,i}$ and the old $(i+1)$-th row and we proceed analogously for $i=1$ to $i=\frac{w-1}{2}$. Finally, the augmented matrix $\Pi$ is transformed into an upper triangular matrix with
 $\Pi_{1,1}=1-\frac{z}{2}$, $\Pi_{2,2}=1-\frac{z}{2}-\frac{z^2}{4}$,  $\Pi_{i,i}= \beta_i / \beta_{i-1}$ for $1<i\leq \frac{w-1}{2}$, and $\Pi_{\frac{w+1}{2},\frac{w+1}{2}}= 1 - \frac{z^2}{2} \beta_{\frac{w-3}{2}}/\beta_{\frac{w-1}{2}}$,
   where ${\beta_i}$ satisfies  the same recurrence relation as shown in Eq.~(\ref{Rec_delta_i}), with the alternate initial conditions $\beta_1=1-\frac{z}{2}$ and $\beta_2=1-\frac{z}{2}-\frac{z^2}{4}$. Thus ~\cite{Richard_Bru01}
\begin{equation}
\label{beta_i}
\beta_i=c_1\times \left(\frac{1+\sqrt{1-z^2}}{2} \right)^i+c_2\times \left(\frac{1-\sqrt{1-z^2}}{2} \right)^i,
\end{equation}
where $c_1=\frac{1}{2}+\frac{\sqrt{1-z^2}}{2(1+z)}$ and $c_2=\frac{1}{2}-\frac{\sqrt{1-z^2}}{2(1+z)}$.
Therefore, if $w$ is odd,
  \begin{equation}
 \label{PGF_RT_O}
 \Phi_{RT}(1,z)=\frac{1}{\Pi_{\frac{w+1}{2},\frac{w+1}{2}}}=\frac{\beta_{\frac{w-1}{2}}}{\beta_{\frac{w-1}{2}}-\frac{z^2}{2}\beta_{\frac{w-3}{2}}}.
 \end{equation}
 Replacing  $\Phi_{RT}(0,z)$ with $\frac{1}{1-z^2}$, and $\Phi_{RT}(1,z)$ with Eqs.~(\ref{PGF_RT_E}) and (\ref{PGF_RT_O}) respectively, we get
 \begin{equation}
 \label{psiz}
 \psi(z)=\left\{ \begin{array}{ll} \frac{\alpha_{\frac{w}{2}}-\frac{z^2}{2}\alpha_{\frac{w}{2}-1}}{(1-z^2)\alpha_{\frac{w}{2}}} &  \textrm{if $w$ is even}\\
   \frac{\beta_{\frac{w-1}{2}}-\frac{z^2}{2}\beta_{\frac{w-3}{2}}}{(1-z^2)\beta_{\frac{w-1}{2}}} &  \textrm{if $w$ is odd} \end{array} \right..
 \end{equation}

In Appendix E and F, the following values of  $\psi(z)$ (and its derivatives) at $z=1$ are used:
\begin{eqnarray}
 \label{a1}
 &&\psi(1)=\frac{w}{2},\\
\label{a2}
&&\left.\frac{d}{d z}\psi(z)\right|_{z=1}=a_1 = \frac{w^3}{12} -\frac{w}{3},\\
\label{a3}
&&\left.\frac{d^2}{d z^2}\psi(z)\right|_{z=1}=a_2= \frac{w^5}{30} -  \frac{w^3}{4} + \frac{7w}{15}.
\end{eqnarray}


\section{Moments for the GFPT}  \label{sec:Moments2}
For the case $G(0)$, we have $\Phi_{GFPT}(0,z)=\frac{z^2+z}{2}$ from which it can be found that $\langle GFPT_0  \rangle=\frac{3}{2}$ and $\langle GFPT_0^2  \rangle=\frac{5}{2}$.
For $t>0$, by calculating the first  order derivative with respect to $z$ on both sides of Eq.~(\ref{R_GFPTm}) and posing  $z=1$,  we find,
\begin{eqnarray}\label{R_First_m_GFPT}
 \langle GFPT_t \rangle&=&\left. \frac{\partial}{\partial z}\Phi_{GFPT}(t,z)\right|_{z=1}\nonumber \\
&=& \langle GFPT_{t-1}\rangle+\frac{2 a_1}{w} \langle FPT_{t-1}\rangle.
\end{eqnarray}
Here $w=u+v$ and $a_1$ is given as in Eq.~(\ref{a2}).

Using Eq.~(\ref{R_First_m_GFPT}) recursively, and replacing the expression for $\langle FPT_{k}\rangle$ with $(uv)^k$ found in Eq.~(\ref {MFPT}),
\begin{eqnarray}\label{First_m_GFPT}
 \langle GFPT_t \rangle  
                   &=&\frac{3}{2}+\left[\frac{(u+v)^2}{6}-\frac{2}{3}\right]\frac{(uv)^t-1}{uv-1}.
\end{eqnarray}
By taking an asymptotic expansion of (\ref{First_m_GFPT}), it can be shown that, for large $t$, $\langle GFPT_{t}\rangle \sim (uv)^t$.

Similarly, we can also calculate the second moment of the GFPT to hub $H_u$, referred to as $\langle GFPT_t^2 \rangle$.
By taking the second order derivative on both sides of Eq.~(\ref{R_GFPTm}) and posing $z=1$, for any $t>0$, 
 \begin{eqnarray}\label{D2_GFPT}
      &&\left. \frac{\partial^2}{\partial z^2}\Phi_{GFPT}(t,z)\right|_{z=1}\nonumber\\
      &=&\frac{2}{w}\left\{\left.\frac{w}{2}\times\frac{\partial^2}{\partial z^2}\Phi_{GFPT}(t-1,z)\right|_{z=1}\right.\nonumber\\
      &+&2a_1\langle GFPT_{t-1}  \rangle\langle FPT_{t-1}  \rangle +a_2(\langle FPT_{t-1}  \rangle)^2\nonumber\\
      &+&\left.a_1\times\left.\frac{\partial^2}{\partial z^2}\Phi_{FPT}(t-1,z)\right|_{z=1}\right\},
  \end{eqnarray}
  where $a_2$ is given as in Eq.~(\ref{a3}).
  Therefore, for any $t>0$,
  \begin{eqnarray}\label{R_Second_m_GFPT}
  \nonumber
  &&   \langle GFPT^2_t\rangle =
      \left. \frac{\partial^2}{\partial z^2}\Phi_{GFPT}(t,z)\right|_{z=1}+\langle GFPT_t \rangle \\
      &=&\langle GFPT^2_{t-1}  \rangle +\frac{2w^4-15w^2+28}{30}\langle FPT_{t-1}  \rangle^2\nonumber\\
      \nonumber
      &+&\frac{w^2-4}{6}\left[2\langle GFPT_{t-1}  \rangle\langle FPT_{t-1}  \rangle+\langle FPT^2_{t-1}  \rangle\right].
  \end{eqnarray}
%
In the previous expression we replace $\langle GFPT_{t-1} \rangle$, $\langle FPT_{t-1}  \rangle$ and $\langle FPT^2_{t-1}  \rangle$ from, respectively, Eqs.~(\ref{First_m_GFPT}), (\ref{MFPT_Em}) and (\ref{M2FPTm}), obtaining, for any $t>0$,
   \begin{equation}\label{R_M2_GFPT}
      \langle GFPT^2_t\rangle
      =\langle GFPT^2_{t-1}  \rangle +k_1(uv)^{2t-2}+k_2 (uv)^{t-1},
  \end{equation}
  where
 \begin{equation} \label{K_1}
   k_1=\frac{(w^2-4)(2w^2+uv-9)}{18(uv-1)}+\frac{2w^4-15w^2+28}{30},
   \end{equation}
    and
   \begin{equation} \label{K_2}
   k_2=\frac{2}{3}(w^2-4)-\frac{(w^2-4)(2w^2+uv-9)}{18(uv-1)}.
   \end{equation}

Using Eq.~(\ref{R_M2_GFPT}) recursively, for any $t>0$,
  \begin{equation}\label{Second_m_GFPT}
      \langle GFPT^2_t\rangle
     =\frac{5}{2}\! +\!k_1\frac{(uv)^{2t}\!-\!1}{(uv)^2-1}\! +\!k_2\frac{(uv)^t\! -\!1}{uv-1}.
  \end{equation}
By taking an asymptotic expansion of (\ref{Second_m_GFPT}), it can be shown that, for large $t$, $\langle GFPT_{t}^2\rangle \sim \langle GFPT_{t}\rangle^2 \sim  (uv)^{2t}$ and that $\textrm{Var}(GFPT_t)\sim (uv)^{2t}$.
%

\section{Moments for the FRT}\label{sec:Moments1}
For the case $G(0)$, we have $\Phi_{FRT}(0,z)=z^2$, from which $\langle FRT_0  \rangle=2$ and $\langle FRT_0^2  \rangle=4$ follow.
%
For any $t>0$, by calculating the first order derivative with respect to $z$ on both sides of Eq.~(\ref{R_FRTm}), posing  $z=1$ and recalling Eq.~(\ref{a1}), we find
\begin{eqnarray}\label{First_m_FRT}
 \langle FRT_t \rangle&=&\left. \frac{\partial}{\partial z}\Phi_{FRT}(t,z)\right|_{z=1}=\psi(1)\times  \langle FRT_{t-1}  \rangle\nonumber \\
&=&\cdots\nonumber \\
&=&[\psi(1)]^t \langle FRT_0  \rangle =2\times\left(\frac{u+v}{2}\right)^t.
\end{eqnarray}

Similarly, we can also calculate the second moment of the FRT to hub $H_0$, referred to as $\langle FRT_t^2 \rangle$.
By taking the  second order derivative on both sides of Eq.~(\ref{R_FRTm}), letting $z=1$ and recalling Eqs.~(\ref{a1}) and (\ref{a2}), we obtain 
 \begin{eqnarray}\label{D2_FRT}
      &&\left. \frac{\partial^2}{\partial z^2}\Phi_{FRT}(t,z)\right|_{z=1} = \left.\psi(1)\times\frac{\partial^2}{\partial z^2}\Phi_{FRT}(t-1,z)\right|_{z=1}\nonumber\\
      &+&2a_1\times\langle FRT_{t-1}  \rangle\times\langle FPT_{t-1}  \rangle.
  \end{eqnarray}
   Therefore, the second moment of FRT for the node $H_0$  satisfies
  \begin{eqnarray}\label{R_Second_m_FRT}
  \nonumber
  \langle FRT^2_t\rangle &=&
      \left. \frac{\partial^2}{\partial z^2}\Phi_{FRT}(t,z)\right|_{z=1}+\langle FRT_t \rangle \\
      \nonumber
      &=&\psi(1)\langle FRT^2_{t-1}  \rangle +2a_1\langle FRT_{t-1}  \rangle\langle FPT_{t-1}  \rangle.
  \end{eqnarray}
Note that $\langle FRT_k \rangle=\psi(1) \langle FRT_{k-1}\rangle$ and $\langle FPT_{k}\rangle=\langle FPT_{1}\rangle^k$. Using Eq.~(\ref{R_Second_m_FRT}) recursively, 
 we get
  \begin{eqnarray}\label{Second_m_FRT}
     \langle FRT^2_t\rangle
     &=&[\psi(1)]^t\langle FRT^2_{0}  \rangle +\frac{2 a_1}{\psi(1)\langle FPT_1  \rangle}\langle FRT_{t}  \rangle \langle FPT_{t}\rangle\nonumber \\
     &\times&\frac{1-\frac{1}{\langle FPT_1  \rangle^t}}{1-\frac{1}{\langle FPT_1  \rangle}}\nonumber \\
     \nonumber
          &=&\left(\frac{u+v}{2}\right)^t  \left \{ \frac{2}{3}[(u+v)^2-4] \frac{(uv)^t-1}{uv-1} + 4 \right \},
  \end{eqnarray}
  which implies that $\langle FRT^2_t\rangle \sim\langle FRT_{t}  \rangle \langle GFPT_{t}\rangle$.


\end{document}